\documentclass[oneside,12pt]{article}
\pdfoutput=1 
\input{setup.sty}

\title{Exploring the holographic entropy cone via reinforcement learning}

\author[a]{Temple He,}
\emailAdd{templehe@caltech.edu}

\author[a]{Jaeha Lee,}
\emailAdd{jaeha@caltech.edu}

 \author[a, b]{Hirosi Ooguri}
\emailAdd{ooguri@caltech.edu}

\affiliation[a]{Walter Burke Institute for Theoretical Physics \rm \& \it Leinweber Forum for Theoretical Physics
\\ California Institute of Technology, Pasadena, CA 91125 USA}

\affiliation[b]{Kavli Institute for the Physics and Mathematics of the Universe \rm (WPI) \\ \it University of Tokyo, Kashiwa 277-8583, Japan}

\abstract{We develop a reinforcement learning algorithm to study the holographic entropy cone. Given a target entropy vector, our algorithm searches for a graph realization whose min-cut entropies match the target vector. If the target vector does not admit such a graph realization, it must lie outside the cone, in which case the algorithm finds a graph whose corresponding entropy vector most nearly approximates the target and allows us to probe the location of the facets. For the $\sf N=3$ cone, we confirm that our algorithm successfully rediscovers monogamy of mutual information beginning with a target vector outside the holographic entropy cone. We then apply the algorithm to the $\sf N=6$ cone, analyzing the 6 {\it mystery} extreme rays of the subadditivity cone from \cite{He:2024xzq} that satisfy all known holographic entropy inequalities yet lacked graph realizations. We found realizations for 3 of them, proving they are genuine extreme rays of the holographic entropy cone, while providing evidence that the remaining 3 are not realizable, implying unknown holographic inequalities exist for $\sf N=6$.}

\begin{document}
\hfill{CALT-TH 2026-005}
 
 
\maketitle

\section{Introduction}

The study of the quantum entropy cone (QEC) is an outstanding research program in the field of quantum information theory, and its goal is to fully specify the set of subsystem entropies realizable by a quantum system involving $\N$ parties for any positive integer $\N$. Subadditivity (SA) and strong subadditivity (SSA) are two such inequalities that constrain the subsystem entropies, and by utilizing the purification symmetry, one can also obtain the Araki-Lieb inequality and weak monotonicity. It is known that for $\N=3$ parties these inequalities are both necessary and sufficient to fully specify the QEC \cite{1193790}. However, it is an open question whether there are further inequalities that are obeyed by all quantum states for larger $\N$.

In this paper, we will be focusing on a subset of the QEC known as the holographic entropy cone (HEC). The HEC is a geometric object introduced in \cite{Bao:2015bfa} that fully characterizes the set of allowed entropy vectors associated to holographic states. The study of the HEC is more tractable due to the fact the entropies of holographic states can be computed using Ryu-Takayanagi (RT) surfaces \cite{Ryu:2006bv}. It was further shown in \cite{Bao:2015bfa} that this calculation can be reduced to finding the min cuts of discrete graphs. As a result, one can utilize properties of discrete graphs to prove inequalities that must be obeyed by such min cuts \cite{He:2019ttu, He:2020xuo, Avis:2021xnz, Czech:2021rxe, Fadel:2021urx, Hernandez-Cuenca:2022pst}, and such inequalities are known as holographic entropy inequalities (HEIs). This line of research has led to the full characterization of the HEC for $\N \leq 5$ \cite{Bao:2015bfa,Hernandez-Cuenca:2019jpv}, and much progress has been made for characterizing the HEC for $\N \geq 6$ \cite{Czech:2022fzb, Hernandez-Cuenca:2023iqh, Czech:2024rco, Bao:2024obe, Bao:2024azn, Bao:2025sjn, Grimaldi:2025jad, Czech:2026tgj, Grimaldi:2026lbq, Czech:2026zca}. Along the way, by utilizing the max flow-min cut theorem, the min cuts can be equivalently thought of as max flows along the edges of the discrete graph, giving an alternative perspective to computing holographic entanglement entropy using ``bit threads'' \cite{Freedman:2016zud, Headrick:2017ucz, Cui:2018dyq, Harper:2019lff, Headrick:2020gyq, Headrick:2022nbe}. Furthermore, there have been various generalizations of the HEC to allow for an entropy cone that is the closure of realizable entropy vectors beyond holographic states \cite{Bao:2020zgx, Bao:2021gzu, Walter:2020zvt, Bao:2020mqq, He:2023cco, He:2023aif, He:2023rox}. 

Despite our improved understanding of the HEC compared to that of the QEC
-- 
and indeed there is now in principle a systematic method of discovering new facets of the HEC \cite{Bao:2024obe, Bao:2024azn, Bao:2025sjn}
--
in practice it is still very challenging to fully characterize the HEC for any $\N \geq 6$ by finding all its facets or all its extreme rays due to the combinatorics involved. As the number of parties $\N$ increases, the number of possible inequalities typically grows doubly exponentially,\footnote{The number of subsystem entropies grow exponentially with respect to $\N$, and each entropy inequality involves a subset of such entropies, resulting in another exponential.} and it quickly becomes computationally intractable to find new facets of extreme rays of the HEC using brute force. As an example, for $\N=3$ and $\N=4$, there is only one class of HEI (up to permutation and purification symmetry), namely the monogamy of mutual information (MMI) \cite{Hayden:2011ag}. For $\N=5$, there are five new classes of HEIs \cite{Bao:2015bfa, Hernandez-Cuenca:2019jpv}. However, for $\N=6$, by utilizing numerous clever techniques, \cite{Hernandez-Cuenca:2023iqh} has already discovered over 1800 new classes of HEIs, and it is expected that there are more as-yet-unknown HEIs.

Due to the challenge described above, a fruitful program that has been pursued in recent years is to focus on the extreme rays of the entropy cone that is carved out by instances of SA alone \cite{Hernandez-Cuenca:2019jpv, Hernandez-Cuenca:2022pst}, and this cone is called the subadditivity cone (SAC). The SAC has many interesting properties that were discovered in \cite{Hernandez-Cuenca:2022pst, He:2022bmi, He:2023cco, He:2023aif, He:2024xzq, Hubeny:2024fjn}, and even though this appears to be a severe restriction, there is a conjecture in \cite{Hernandez-Cuenca:2022pst} stating that all the extreme rays of the HEC involving $\N$ parties can be determined from a subset of extreme rays of an SAC involving $\N' \geq \N$ parties. If this is indeed true, a full understanding of the SAC at any $\N'$ will then offer invaluable information pertaining to HECs at lower party numbers. Even if the conjecture of \cite{Hernandez-Cuenca:2022pst} turns out to be false, we can still find new extreme rays of the HEC by first determining the extreme rays of the SAC and then ascertaining which ones are associated to min cuts of discrete graphs.\footnote{Note that any extreme ray of the SAC that is inside the HEC must also be an extreme ray of the HEC. This follows because the HEC lies inside the SAC.} Moreover, the SAC is an outer bound for the QEC, while the HEC is an inner bound. Therefore, the SAC and HEC together are able to provide both outer and inner bounds to constrain the QEC. 

Our goal in this paper is to introduce a machine learning (ML) algorithm to tackle some of the problems described above. Reinforcement learning (RL) has proven to be an effective framework for exploring high-dimensional search spaces, with notable applications including game playing \cite{Silver:2016AlphaGo}, protein structure prediction \cite{Jumper:2021AlphaFold}, and the discovery of novel mathematical algorithms \cite{Fawzi:2022AlphaTensor}. In the context of theoretical physics and mathematics, RL has been utilized to explore the landscape of string vacua \cite{Halverson:2019branes} and to discover sequences of moves that simplify knots \cite{Gukov:2021unknot}, in both cases significantly outperforming random search methods. The primary advantage of RL lies in its ability to efficiently navigate large combinatorial spaces by learning to preferentially sample regions that yield higher rewards, making it well-suited to problems where the search space grows exponentially and the
  objective function is non-smooth.

The usage of ML in studying the HEC has surprisingly been relatively uncommon.\footnote{Recently, \cite{Khumalo:2025xfv} did use reinforcement learning to explore properties of the entropy cone. Furthermore, \cite{VanRaamsdonk:2025knt} also used artificial intelligence to explore properties of entanglement entropy.} Here, we hope to demonstrate that even with a simple vanilla policy gradient algorithm, we can discover new aspects of the HEC. Given a target vector $\vec\ent$ in entropy space, our RL algorithm attempts to find a graph whose min-cut entropies match the target. The policy network takes the current graph configuration as input and outputs updated edge weights, with a reward function defined as the cosine similarity between the achieved and target entropy vectors. If the target lies inside the HEC, the algorithm finds an exact realization and the reward reaches its maximum value of $1$; if the target lies outside, the algorithm finds the graph
that approximates the target as closely as possible. As with all ML procedures, the output has finite numerical precision. However, we can compute the entropy vector analytically from the graph outputted by the algorithm, allowing us to verify that the realization is exact and not an artifact of numerical rounding errors.

A key feature of our approach is that the reward function serves a dual purpose. First, the optimized reward directly indicates classification: an entropy vector achieving the maximum reward of 1 lies inside the HEC, while one with a lower maximal reward lies outside. Second, when the target entropy vector lies outside the cone, the gradient of the reward function points towards the nearest boundary, effectively navigating the agent towards potentially undiscovered facets. We demonstrate both capabilities systematically. As a proof of concept, we apply our algorithm to the $\N=3$ case, where the HEC is completely specified by SA and MMI, and the reward landscape can be computed analytically. This allows us to validate that our RL-trained agents correctly classify extreme rays and that the learned gradient flow aligns with the analytical prediction. In particular, starting from an extreme ray of the SAC that violates MMI, we show that the gradient accurately points towards the MMI boundary, successfully rediscovering this holographic inequality. We then turn to $\N=6$, where analytical treatment is no longer feasible and RL becomes essential. It was determined in \cite{He:2024xzq} that there are 208 new classes of extreme rays of the SAC satisfying SSA for $\N=6$. Of these 208 rays, there are 6 which are dubbed ``mystery rays.'' These 6 extreme rays satisfy every known holographic entropy inequality for $\N=6$, and yet the authors of \cite{He:2024xzq} did not find a graph realization for these 6 rays. Using our algorithm, we determine that 3 of these 6 mystery rays have graph realizations and are therefore extreme rays of the $\N=6$ HEC as well. As for the other 3, our results suggest they are likely not realizable.

Our target audience for this paper is for physicists interested in studying the HEC who may not be as familiar with RL techniques, and therefore we will devote more effort to explain the RL concepts involved. Our outline is as follows. In \Cref{sec:setup}, we will give a brief review of the holographic entropy cone and also present the relevant background in RL. In \Cref{sec:mmi}, we present a proof of concept by applying our algorithm to the $\N=3$ case, where we validate the classification capability and demonstrate the gradient-based rediscovery of MMI. In \Cref{sec:realizability}, we apply our algorithm to the $\N=6$ SAC, potentially classifying all 208 extreme rays and resolving the status of the mystery rays. We conclude with some discussions and future directions in \Cref{sec:discussion}. Our code implementing our RL algorithm can be found at \href{https://github.com/Jaeha0526/EntropyCone_RL}{https://github.com/Jaeha0526/EntropyCone\_RL}.

\section{Setup}\label{sec:setup}

We begin by introducing the necessary prerequisites. In \Cref{ssec:hec}, we will review of the holographic entropy cone (HEC) and standardize the notation we will use. In \Cref{ssec:ml-alg}, we will review the conceptual basis of reinforcement learning, as well as introduce the algorithm we will utilize in our exploration of the HEC.

\subsection{Holographic entropy cone}\label{ssec:hec}

Consider a quantum system with $\N$ parties, labeled by $1, 2,\ldots, \N$.\footnote{For small $\N$, it is also customary to label the parties as $A,B,C,\ldots$.} We can study the entanglement structure of this system by specifying the von Neumann entropy associated to nonempty subsets of the parties, i.e. $\ent(I)$ for $I \subseteq [\N] := \{1,2,\ldots,\N\}$. There are $2^{\N}-1$ such entropies, and therefore we can view the subsystem entropies as components of a vector living in a vector space with $\D = 2^\N-1$ dimensions. Such a vector is called an entropy vector, and the vector space is called the entropy space.

It is clear that not every vector in the entropy space is realizable by a quantum state. For instance, we are restricted to the region of entropy space where the subsystem entropies are all nonnegative, in that
\begin{align}
    \ent(I) \geq 0 \quad\text{for all $I \subseteq [\N]$},
\end{align}
since the von Neumann entropy is always nonnegative. A natural question is to ask what are the vectors in the entropy space that is realizable by a quantum state. It was shown in \cite{1193790} that the topological closure of all such realizable entropy vectors form a convex cone, and we refer to this cone as the quantum entropy cone (QEC). For $\N = 3$, we can fully characterize the associated QEC. Indeed, we know that entropies of quantum states have to obey four classes of inequalities, which we list below:
\begin{align}
    \text{Subadditivity (SA):} \quad & \ent(I) + \ent(J) \geq \ent(IJ) \\
    \text{Araki-Lieb (AL):} \quad & \ent(I) + \ent(IJ) \geq \ent(J) \\
    \text{Strong subadditivity (SSA):} \quad & \ent(IJ) + \ent(JK) \geq \ent(J) + \ent(IJK) \\
    \text{Weak monotonicity (WM):} \quad & \ent(IJ) + \ent(JK) \geq \ent(I) + \ent(K) ,
\end{align}
where we use the shorthand $IJ := I \cup J$. These four classes of inequalities hold for every possible disjoint subset of parties $I,J,K$ and constrain the set of allowable entropy vectors that are realizable by quantum states to be a convex cone. Furthermore, it is also known that every entropy vector within the cone constrained by the inequalities above lies within the topological closure of a realizable entropy vector \cite{1193790}, meaning that the QEC for $\N=3$ is completely specified by the above inequalities. However, for $\N \geq 4$, fully specifying the QEC in terms of a set of inequalities is an open question.

Thus far, we have not utilized the fact that we can purify our system. Consider now adding a purifier party  $\N+1$,\footnote{The purifier party is also sometimes denoted $O$.} with the condition that
\begin{align}
    \ent(1\cdots\N+1) = 0 .
\end{align}
We denote $\uI,\uJ,\ldots$ to be subsets of $[\N+1]$, which includes the purifier, and we define $\uI^c := [\N+1] \setminus \uI$ to be the complement of $\uI$. Purification symmetry is then the property that
\begin{align}\label{purification-sym}
    \ent(\uI) = \ent(\uI^c).
\end{align}
We can now use \eqref{purification-sym} to verify that the four classes of inequality above can actually be simplified down to two inequalities, namely
\begin{align}
    \text{Subadditivity (SA):} \quad & \ent(\uI) + \ent(\uJ) \geq \ent(\uI\uJ) \label{sa} \\
    \text{Strong subadditivity (SSA):} \quad & \ent(\uI\uJ) + \ent(\uJ\uK) \geq \ent(\uJ) + \ent(\uI\uJ\uK), \label{ssa}
\end{align}
where we have now included the purifier. Indeed, AL and WM are related to SA and SSA via purification, respectively.

Due to the challenges facing the complete characterization of the QEC for $\N \geq 4$, we instead focus exclusively on holographic states, which are the subset of quantum states that admit a geometric holographic dual and have entropies that can be computed using Ryu-Takayanagi (RT) surfaces \cite{Ryu:2006bv}. The topological closure of entropy vectors realizable by holographic states is the holographic entropy cone (HEC), and it is a convex, polyhedral cone that is within the QEC \cite{Bao:2015bfa, Avis:2021xnz}. We know that the HEC is strictly smaller than the QEC because holographic states obey additional entropy inequalities that could be violated by quantum states. Such inequalities are known as holographic entropy inequalities (HEIs), and they form the facets of the HEC. The most well-known example of such an inequality is the monogamy of mutual information (MMI), namely
\begin{align}\label{mmi}
\begin{split}
    &\text{Monogamy of mutual information (MMI):} \\
    &\qquad\qquad \ent(\uI\uJ) + \ent(\uI\uK) + \ent(\uJ\uK) \geq \ent(\uI) + \ent(\uJ) + \ent(\uK) + \ent(\uI\uJ\uK).
\end{split}
\end{align}
It has been proven in \cite{Hayden:2011ag, Cui:2018dyq} that \eqref{mmi} is satisfied for all holographic states. On the other hand, it is simple to find a quantum state that violates it. Consider the Greenberger-Horne-Zeilinger (GHZ) state for $\N=3$, which has the subsystem entropies \cite{Greenberger:1989tfe}
\begin{align}\label{ghz3}
    \ent(A) = \ent(B) = \ent(C) = \ent(AB) = \ent(AC) = \ent(BC) = \ent(ABC) = 1.
\end{align}
It follows immediately that MMI is violated, implying that the GHZ state is not holographic. Indeed, one can view MMI as a stronger version of SSA, since MMI and SA together imply SSA.

There has been more progress in the exploration of the HEC than that of the QEC. We know what the HEC is up to $\N=5$ \cite{Bao:2015bfa,Hernandez-Cuenca:2019jpv}, and many facets of the $\N=6$ HEC have been uncovered in recent years \cite{Hernandez-Cuenca:2023iqh}. The primary reason why we are able to make such progress is due to the fact that unlike entropy vectors of generic quantum states, we can represent holographic entropy vectors in terms of a discrete graph \cite{Bao:2015bfa}. More concretely, given a holographic entropy vector $\vec\ent$, we can associate to it a graph $\CG = (V,E)$ with vertices $V$ and edges $E$ such that a subset of the vertices $\p V := [\N+1]$, which we call boundary vertices, represent the parties (including the purifier), and each edge carries a weight. We further denote the remaining vertices $V \setminus \p V$ to be internal vertices. The entropy associated to any subsystem of boundary vertices $\uI$ is then given by the min cut that separates $\uI$ (in addition to possible internal vertices) from its complement $\uI^c := \p V \setminus \uI$.\footnote{Alternatively, it was shown in \cite{Freedman:2016zud} that instead of min cuts, one can also work with max flows.} Because of this graph representation of the entropy vector, \cite{Bao:2015bfa} demonstrated that one may use what are known as contraction maps to prove new HEIs. In fact, it was recently shown that every single HEI arises from a contraction map \cite{Bao:2024obe, Bao:2024azn, Bao:2025sjn}. This means that we have a conceptual framework for how to find new HEIs, and that the main bottleneck for such exploration is the large combinatorics involved for even modest number of parties such as $\N=6$. For this reason, the exact specification of the HEC is still a challenging task, and we now turn to introducing the machine learning algorithm we developed to partly help with this task in the next subsection.

\subsection{Reinforcement learning algorithm}\label{ssec:ml-alg}

In this subsection, we explain the conceptual basis of reinforcement learning (RL) in a manner tailored to the problem studied in this paper. Our aim is to provide theoretical physicists with sufficient intuition to understand why RL is a natural and effective tool for exploring the HEC, without attempting a general or comprehensive introduction to RL. We therefore focus exclusively on aspects of RL that are directly relevant to the task of determining whether a given entropy vector admits a graph realization.

At its core, reinforcement learning is a framework for solving optimization problems in which one seeks to maximize a reward function through a sequence of adaptive decisions. Unlike supervised learning, which relies on a preexisting dataset of labeled examples, RL proceeds by repeatedly proposing candidate solutions, evaluating their quality via a reward, and updating future proposals accordingly. This paradigm is particularly well-suited to the present setting, where the search space is highly combinatorial, the objective function is non-smooth, and there is no natural training set.

\paragraph{Entropy realizability as a search problem.}
The central computational task in this paper is the following. Given a target entropy vector $\vec\ent_{\rm target}$ in the entropy space of $\N$ parties (including a purifier), we would like to determine whether there exists a weighted graph $\CG$ whose min cuts reproduce $\vec\ent_{\rm target}$. If such a graph exists, the vector lies in the HEC; if not, it does not.

A graph realization consists of $\N+1$ boundary vertices (representing the $\N$ parties plus the purifier) together with $n$ internal vertices, for a total of $n_T := \N+1+n$ vertices. Adding more internal vertices increases the realizability power of the graph: it is known that for $\N = 1, 2, 3, 4$, a total of $2, 3, 5, 6$ vertices respectively suffice to realize all entropy vectors inside the HEC \cite{Avis:2021xnz}. In our RL approach, we fix the number of internal vertices $n$ and search over edge weights of a complete graph on all $n_T$ vertices, restricting weights to be non-negative. For the $\N=3$ validation in \Cref{sec:mmi}, we use $n=1$ internal vertex (5 vertices total), which is known to be sufficient. For the $\N=6$ classification in \Cref{sec:realizability}, we explore graphs with up to $n=13$ internal vertices (20 vertices total).

From the RL viewpoint, this question is naturally formulated as a search problem. One begins with an initial graph ansatz and iteratively adjusts the edge weights to reduce the discrepancy between the entropy vector $\vec\ent(\CG)$ induced by the graph and the target vector $\vec\ent_{\rm target}$. Each modification is evaluated, and information gained from this evaluation is used to guide subsequent modifications. RL provides a principled framework for organizing and optimizing this type of adaptive search. However, since the action space grows as $O(n_T^2)$ with the number of vertices, finding the optimal edge weight configuration becomes increasingly challenging as $n_T$ increases, even with RL's efficient exploration.

\paragraph{States, actions, and environment.}
In RL terminology, the \emph{state} encodes the current candidate graph together with its edge weights. The \emph{action space} consists of allowed updates of this state, such as adjusting edge weights or modifying the internal connectivity of the graph while keeping the boundary vertices fixed. The \emph{environment} is the deterministic map that takes a proposed graph to its associated entropy vector by computing the relevant min cuts, and then evaluates a reward based on how close this vector is to $\vec\ent_{\rm target}$.

The core of RL is a neural network called the \emph{policy}, which learns to map states to actions. Given a state, the policy outputs a probability distribution over possible actions, and an action is sampled from this distribution. The training objective is to adjust the network parameters so that the policy selects actions leading to high expected cumulative reward. Through repeated interaction with the environment -- proposing actions, observing rewards, and updating the policy -- the network gradually learns which modifications to the graph are most likely to improve the reward.

An important feature of this environment is that it is highly non-smooth. The entropy $\ent(I)$ is given by a minimum over cuts, so small changes in edge weights may leave all entropies unchanged until a critical threshold is crossed, at which point the minimizing cut changes discontinuously. Furthermore, changes in graph topology are intrinsically discrete. For these reasons, standard gradient-based optimization techniques are poorly suited to this problem: they require differentiating the reward function with respect to the graph parameters, which is ill-defined when the reward landscape has discontinuities.

RL circumvents this difficulty by never requiring gradients of the reward function itself. Instead, RL computes gradients with respect to the \emph{policy parameters} (i.e., the neural network weights), using only the numerical values of rewards obtained from sampled actions. Concretely, the algorithm samples many actions from the current policy, observes which actions lead to higher rewards, and updates the policy to increase the probability of high-reward actions. This approach treats the environment as a black box and remains valid regardless of whether the reward landscape is smooth, discontinuous, or even stochastic.

\paragraph{Reward function and physical meaning.} 
The reward function is where the physics problem is encoded. In our application, we define the reward as the cosine similarity between the target entropy vector and the entropy vector induced by the graph:
\begin{align}\label{eq:reward-def}
    R = \frac{\vec\ent_{\rm target} \cdot \vec\ent(\CG)}{\|\vec\ent_{\rm target}\| \, \|\vec\ent(\CG)\|}.
\end{align}
Since entropy is non-negative, all components of both vectors lie in the positive orthant, so the reward ranges from $0$ to $1$. The maximum $R = 1$ is achieved if and only if the two vectors are parallel, i.e., $\vec\ent(\CG) = \lambda \vec\ent_{\rm target}$ for some $\lambda > 0$. Since the HEC is a cone, positive scalar multiples of realizable vectors are also realizable, so achieving $R = 1$ certifies that the target lies inside the HEC. Conversely, if the target lies outside the HEC, no graph can achieve $R = 1$, and the maximum achievable reward indicates how close the target is to the cone boundary. In this way, the abstract RL objective of maximizing reward is directly aligned with the physical goal of finding a graph realization.

This formulation also provides a useful bridge between numerical exploration and exact results. Although the RL algorithm operates with finite numerical precision, any promising output graph can be analyzed independently and exactly: its min cuts can be recomputed analytically, and its edge weights can often be rescaled to rational or integer values. Thus, RL serves as a discovery tool that proposes candidates, while final certification of realizability remains fully analytic.

\section{Proof of concept: the $\N=3$ HEC}\label{sec:mmi}

In this section, we use the $\N=3$ case as a proof of concept for our RL algorithm. The $\N=3$ HEC is sufficiently simple that the reward landscape can be computed analytically, allowing us to validate that our RL algorithm correctly recovers both the classification capability (reward $R = 1$ inside the cone) and the gradient-based navigation towards cone boundaries. We will demonstrate that starting from a point outside the HEC, the RL-computed gradient accurately points towards the MMI facet, successfully rediscovering this holographic inequality.

\subsection{Review of the $\N=3$ HEC}\label{ssec:n3-review}

Consider the $\N=3$ holographic entropy cone with parties $A,B,C$ and purifier $O$. The entropy vectors are 7-dimensional, and can be labeled in the following lexicographic manner:\footnote{For arbitrary $\N$, lexicographic ordering means we first write the single-party entropies in alphabetical (or numerical) ordering, followed by the 2-party entropies $AB, AC, \ldots, BC, BD, \ldots$, and so forth.}
\begin{align}
    \vec\ent = \big\{ \ent(A) , \ent(B) , \ent(C) ; \ent(AB) , \ent(AC) , \ent(BC) ; \ent(ABC) \big\}.
\end{align}
The only HEI for $\N=3$ is MMI given in \eqref{mmi}, which in this case is simply
\begin{align}\label{eq:mmi-n3}
    \ent(AB) + \ent(AC) + \ent(BC) \geq \ent(A) + \ent(B) + \ent(C) + \ent(ABC).
\end{align}
Together with SA, these inequalities form the facets of the $\N=3$ HEC. Note that SSA is a positive linear combination of MMI and SA, making it a redundant constraint that can be neglected.

To demonstrate our RL algorithm's effectiveness, suppose we only know the SAs for $\N=3$ and would like to rediscover MMI. By imposing all 18 instances of SA, we obtain the $\N=3$ SAC. This cone consists of 11 extreme rays (ignoring permutation and purification symmetry): 7 of them lie inside the HEC (satisfying MMI), namely the 6 Bell pairs and the 4-party pefect tensor, and 4 lie outside (violating MMI). Among the 4 MMI-violating extreme rays, the one that violates MMI the most is given by (normalized so that the sum of components equals 1)
\begin{align}\label{eq:sac-extreme-vec}
    \vec\ent_{\text{SAC}} := \frac{1}{4} \{1,1,1;0,0,0;1\} .
\end{align}
Our goal is to show that starting from this vector, the RL algorithm can navigate towards the HEC boundary and rediscover MMI by following the gradient of the reward function. However, before pursuing this goal, we need to understand the behavior of the reward function outside the HEC, which we now turn to.

\subsection{Analytical reward landscape}\label{ssec:analytical-landscape}

To demonstrate the utility of the reward landscape as a tool for understanding the HEC, we derive closed-form expressions on the $S_3$-symmetric slice of the $\N=3$ entropy space. For $\N=3$, it is known that graphs with 5 vertices suffice to realize all interior points of the HEC \cite{Avis:2021xnz}: 4 boundary vertices (parties $A$, $B$, $C$ and purifier $O$) and 1 internal vertex $X$. Since the HEC is a cone, if $\vec\ent$ is realizable then so is $\lambda\vec\ent$ for any $\lambda > 0$. We therefore work on the unit sphere $\|\vec\ent\|^2 = 1$ without loss of generality, which reduces the dimensionality by one. This analytical treatment reveals how the reward function encodes the cone structure, and in \Cref{ssec:rl-validation} we will verify that our RL algorithm successfully recovers these theoretical bounds.

\begin{figure}[tb]
    \centering
    \includegraphics[width=0.75\textwidth]{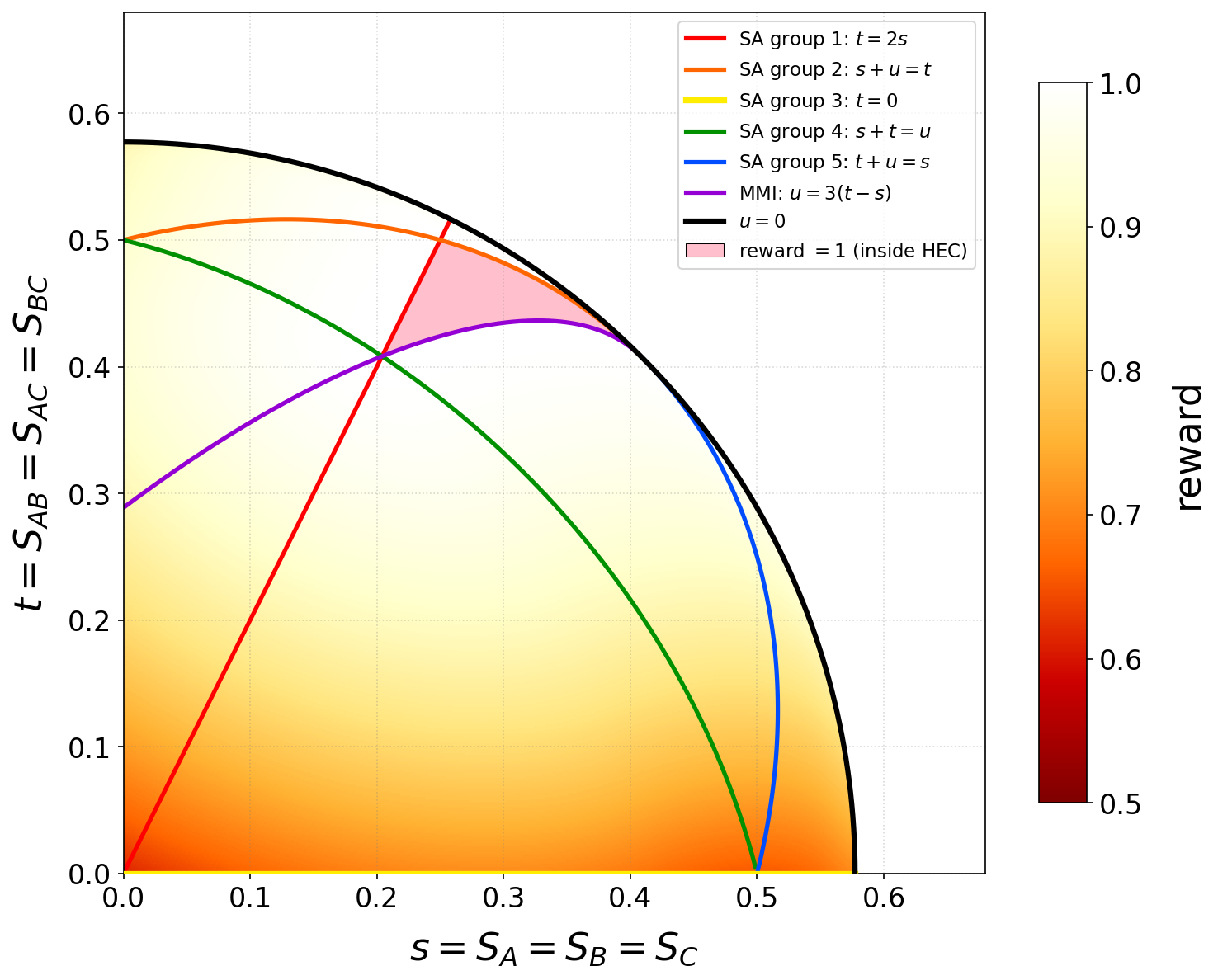}
    \caption{The $S_3$-symmetric slice of the $\N=3$ entropy space in the $(s,t)$ plane. The color gradient shows the analytically computed reward $R$ (cosine similarity). On this symmetric slice, the HEC (pink region, $R = 1$) is constrained by MMI, SA group 1, and SA group 2. The black circle is the boundary $u = 0$.}
    \label{fig:st-regions}
\end{figure}

\paragraph{Symmetric parameterization.}
We further restrict to the $S_3$-symmetric subspace where all parties are equivalent:
\begin{align}
    \vec\ent = \{s, s, s; t, t, t; u\},
\end{align}
with $s = \ent(A) = \ent(B) = \ent(C)$, $t = \ent(AB) = \ent(AC) = \ent(BC)$, and $u = \ent(ABC)$. The unit sphere constraint $3s^2 + 3t^2 + u^2 = 1$ determines $u$ in terms of $(s,t)$:
\begin{align}\label{eq:u-constraint}
    u = \sqrt{1 - 3s^2 - 3t^2}.
\end{align}
This reduces the 7-dimensional entropy space to a 2-dimensional disk of radius $1/\sqrt{3}$ in the $(s,t)$ plane, as shown in \Cref{fig:st-regions}.

\paragraph{HEC constraints on the symmetric slice.}
The $\N=3$ SAC is defined by 18 SA inequalities, arising from 6 triples of subsets each generating 3 triangle inequalities. On the symmetric slice, these 18 inequalities reduce to 5 distinct groups as shown in \Cref{tab:sa-groups}. The HEC is further constrained by MMI, namely $u \leq 3(t-s)$, and is shown in pink in \Cref{fig:st-regions}.

\begin{table}[tb]
    \centering
    \begin{tabular}{c p{8.5cm} l}
        \hline
        Group & Inequalities & Symmetric form \\
        \hline
        1 & $\ent(A) + \ent(B) \geq \ent(AB)$, $\ent(A) + \ent(C) \geq \ent(AC)$, $\ent(B) + \ent(C) \geq \ent(BC)$ & $t \leq 2s$ \\[6pt]
        2 & $\ent(A) + \ent(ABC) \geq \ent(BC)$, $\ent(B) + \ent(ABC) \geq \ent(AC)$, $\ent(C) + \ent(ABC) \geq \ent(AB)$ & $t \leq s + u$ \\[6pt]
        3 & $\ent(A) + \ent(AB) \geq \ent(B)$, $\ent(B) + \ent(AB) \geq \ent(A)$, $\ent(A) + \ent(AC) \geq \ent(C)$, $\ent(C) + \ent(AC) \geq \ent(A)$, $\ent(B) + \ent(BC) \geq \ent(C)$, $\ent(C) + \ent(BC) \geq \ent(B)$ & $t \geq 0$ \\[6pt]
        4 & $\ent(A) + \ent(BC) \geq \ent(ABC)$, $\ent(B) + \ent(AC) \geq \ent(ABC)$, $\ent(C) + \ent(AB) \geq \ent(ABC)$ & $u \leq s + t$ \\[6pt]
        5 & $\ent(BC) + \ent(ABC) \geq \ent(A)$, $\ent(AC) + \ent(ABC) \geq \ent(B)$, $\ent(AB) + \ent(ABC) \geq \ent(C)$ & $s \leq t + u$ \\
        \hline
    \end{tabular}
    \caption{The 18 SA inequalities reduce to 5 groups on the symmetric slice.}
    \label{tab:sa-groups}
\end{table}

\paragraph{Analytical reward formula.}
Recall from \Cref{ssec:ml-alg} that our reward $R$ is the cosine similarity between target and achieved entropy vectors. For unit vectors, this is simply the dot product
\begin{align}\label{eq:reward-formula}
    R = \vec\ent_{\text{target}} \cdot \vec\ent_{\text{achieved}} = 3ss' + 3tt' + uu',
\end{align}
where $(s',t',u')$ denotes the achieved entropy vector. If the target lies inside the HEC, the optimal achieved vector is the target itself, giving $R = 1$. If the target lies outside the HEC, the optimal achieved vector is the closest point on the HEC boundary (in the sense of maximizing cosine similarity), giving $R < 1$.

\paragraph{SA group 1 violated region ($t > 2s$).}
When the target violates SA group 1 (subadditivity), the optimal achieved point lies on the boundary $t' = 2s'$. Using Lagrange multipliers (see Appendix~\ref{app:case2}), the optimal solution is
\begin{align}\label{eq:SA-projection}
    s' = \frac{s + 2t}{D_1}, \quad t' = 2s', \quad u' = \frac{5u}{D_1}, \quad \text{where } D_1 = \sqrt{15(s+2t)^2 + 25u^2}.
\end{align}

\paragraph{SA group 2 violated region ($t > s + u$, with $t \leq 2s$).}
When the target violates SA group 2 but satisfies SA group 1, the optimal achieved point lies on the boundary $t' = s' + u'$. Using Lagrange multipliers (see Appendix~\ref{app:case2b}), define
\begin{align}\label{eq:SA2-aux}
    P = 4s + t - u, \quad Q = 3(t-s) + 2u, \quad D = \sqrt{6P^2 + 6P Q + 4Q^2}.
\end{align}
The optimal solution is
\begin{align}\label{eq:SA2-projection}
    s' = \frac{P}{D}, \quad t' = \frac{P + Q}{D}, \quad u' = \frac{Q}{D}.
\end{align}

\paragraph{MMI-violated region ($u > 3(t-s)$, with $t \leq 2s$ and $t \leq s+u$).}
When the target violates MMI but satisfies SA groups 1 and 2, the optimal achieved point lies on the MMI boundary $u' = 3(t' - s')$. The optimal ratio $\rho = t'/s'$ is given by
\begin{align}\label{eq:optimal-rho}
    \rho = \frac{3s + 4t + u}{4s + 3t - u}.
\end{align}
This ratio must satisfy $1 \leq \rho \leq 2$: the lower bound ensures non-negative $u'$, while the upper bound comes from SA group 1. If the computed $\rho$ falls below 1, we set $\rho = 1$; if it exceeds 2, we set $\rho = 2$. The optimal $(s', t', u')$ can then be computed explicitly from $\rho$ to be (see Appendix~\ref{app:case3})
\begin{align}
\begin{split}
    s' = \frac{1}{\sqrt{6(2\rho^2 - 3\rho + 2)}}, \quad t' = \rho s', \quad u' = 3(\rho-1)s'.
\end{split}
\end{align}

\begin{figure}[tb]
    \centering
    \includegraphics[width=0.95\textwidth]{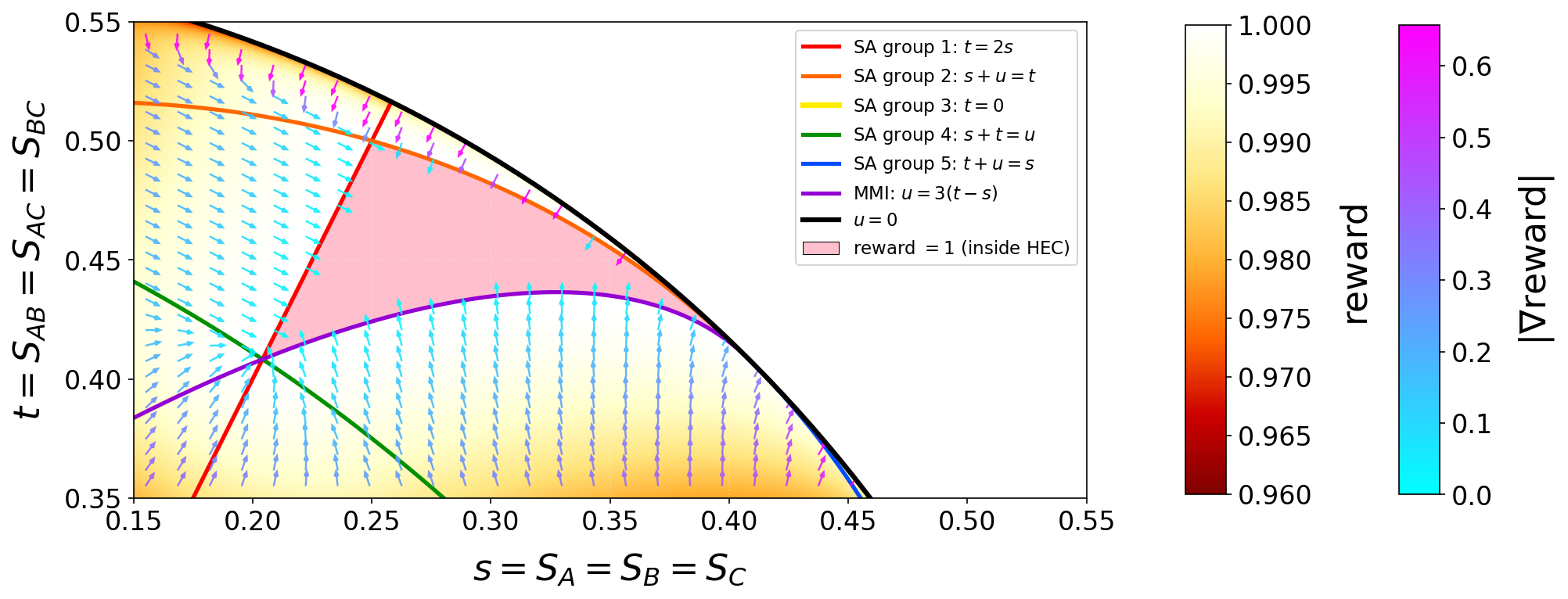}
    \caption{Zoomed in view of the reward landscape near the HEC boundary, showing the gradient field in the region $s \in [0.15, 0.55]$ and $t \in [0.35, 0.55]$. Arrows indicate the direction of steepest reward increase, colored by magnitude. The gradient consistently points towards the nearest HEC boundary (SA group 1, SA group 2, or MMI).}
    \label{fig:gradient-landscape}
\end{figure}

\paragraph{Reward gradient.}
The reward function encodes not only the proximity to the HEC, but also the direction towards it. For points outside the HEC, the gradient of the reward with respect to the target position points towards the nearest HEC boundary. In the $(s,t)$ parameterization,
\begin{align}\label{eq:gradient-formula}
    \frac{\partial R}{\partial s} = 3s' - \frac{3su'}{u}, \quad \frac{\partial R}{\partial t} = 3t' - \frac{3tu'}{u}.
\end{align}
This gradient exhibits several important properties that we expect to generalize beyond the symmetric slice to the full entropy space and higher $\N$. The reward $R$ is maximized inside the HEC, where $R = 1$, and decreases monotonically as we move away from the cone boundary. The gradient consistently points towards the nearest HEC facet, providing a natural ``navigator'' towards undiscovered constraints. Furthermore, as a point approaches a facet boundary, the gradient becomes increasingly aligned with that facet's normal vector; at the boundary itself, the gradient is parallel to the facet normal.
\Cref{fig:gradient-landscape} visualizes this behavior, with arrows indicating the gradient direction (colored by magnitude) and showing how the gradient field consistently points towards the HEC from all exterior regions. These properties suggest using the reward function as a navigator for flowing towards the HEC boundary: by following the gradient from any exterior point, we can reach the cone surface. Moreover, since the gradient aligns with the facet normal near the boundary, reading the gradient direction at the surface allows us to identify the constraining facet---potentially discovering new, previously unknown inequalities. In \Cref{ssec:mmi-rediscovery}, we demonstrate this navigator property explicitly by following the gradient from outside the HEC to rediscover the MMI facet.

\subsection{RL validation of analytical predictions}\label{ssec:rl-validation}

To validate that our RL algorithm correctly recovers the analytical reward landscape, we sample a grid of points across the symmetric slice and compare the RL-achieved rewards against the analytical predictions.

\subsubsection{RL training setup}

For the $\N=3$ validation, we use a graph with 5 vertices: 4 boundary vertices (parties $A$, $B$, $C$ and purifier $O$) and 1 internal vertex $X$. The policy network is a 2-layer feedforward network with 64 hidden units per layer. Training uses a batch size of 60, learning rate $10^{-4}$, rollout length of 50 steps, and a maximum of 2000 iterations per run. We apply exploration noise with standard deviation $0.15$ and perturbation size $\delta t = 0.1$ for gradient estimation. Early stopping with patience 5 is enabled to terminate training once convergence is achieved. For each grid point, we perform 20 independent training runs to obtain robust statistics.

\subsubsection{Grid validation}

\paragraph{Method.}
We sample a $20 \times 20$ grid of points $(s_i, t_j)$ spanning the valid region of the symmetric slice. For each target entropy vector $(s_i, s_i, s_i, t_j, t_j, t_j, u_{ij})$, with $u_{ij} = (1 - 3s_i^2 - 3t_j^2)^{1/2}$, we run our RL algorithm to find the optimal graph realization. We then compare the RL-achieved reward with the analytically computed reward from \eqref{eq:reward-formula}.

\paragraph{Results.}
In our sample of a $20 \times 20$ grid across the symmetric slice, we find that 324 points fall within the valid region (satisfying $3s^2 + 3t^2 < 1$). For each valid point, we run 20 independent training runs and record the maximum reward achieved.

\begin{figure}[tb]
    \centering
    \includegraphics[width=0.7\textwidth]{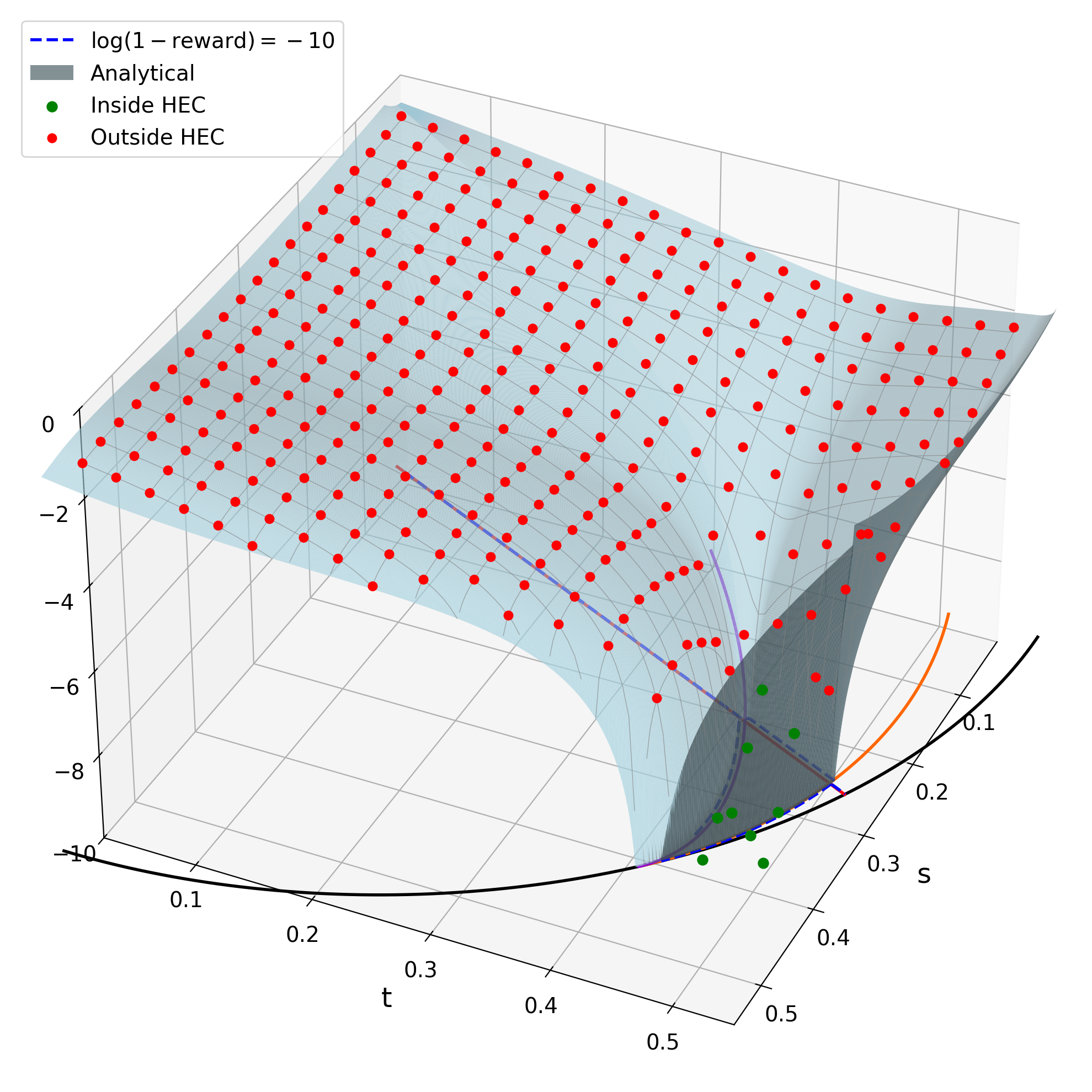}
    \caption{3D view of the reward landscape on the $\N=3$ symmetric slice, showing agreement between analytical predictions (surface) and RL results (points). Rewards are plotted as $\log(1-R)$ for visibility near the HEC boundary. The red dots lie outside the HEC, whereas the green dots lie inside the HEC. }
    \label{fig:rl-3d}
\end{figure}

The RL-achieved rewards closely follow the analytical reward surface. Across all 324 grid points, we observe a Pearson correlation of $0.996$ between the RL results and the analytical predictions, demonstrating that our algorithm accurately captures the reward landscape. \Cref{fig:rl-3d} visualizes this agreement: the analytical surface (colored mesh with grid lines) is overlaid with RL data points, showing excellent correspondence across the entire domain. The funnel-like structure near the HEC boundary region reflects the logarithmic transformation $\log(1-R)$ used for visualization, which makes small differences visible near $R=1$.

\subsubsection{Classification ability}

\begin{figure}[tb]
    \centering
    \includegraphics[width=0.8\textwidth]{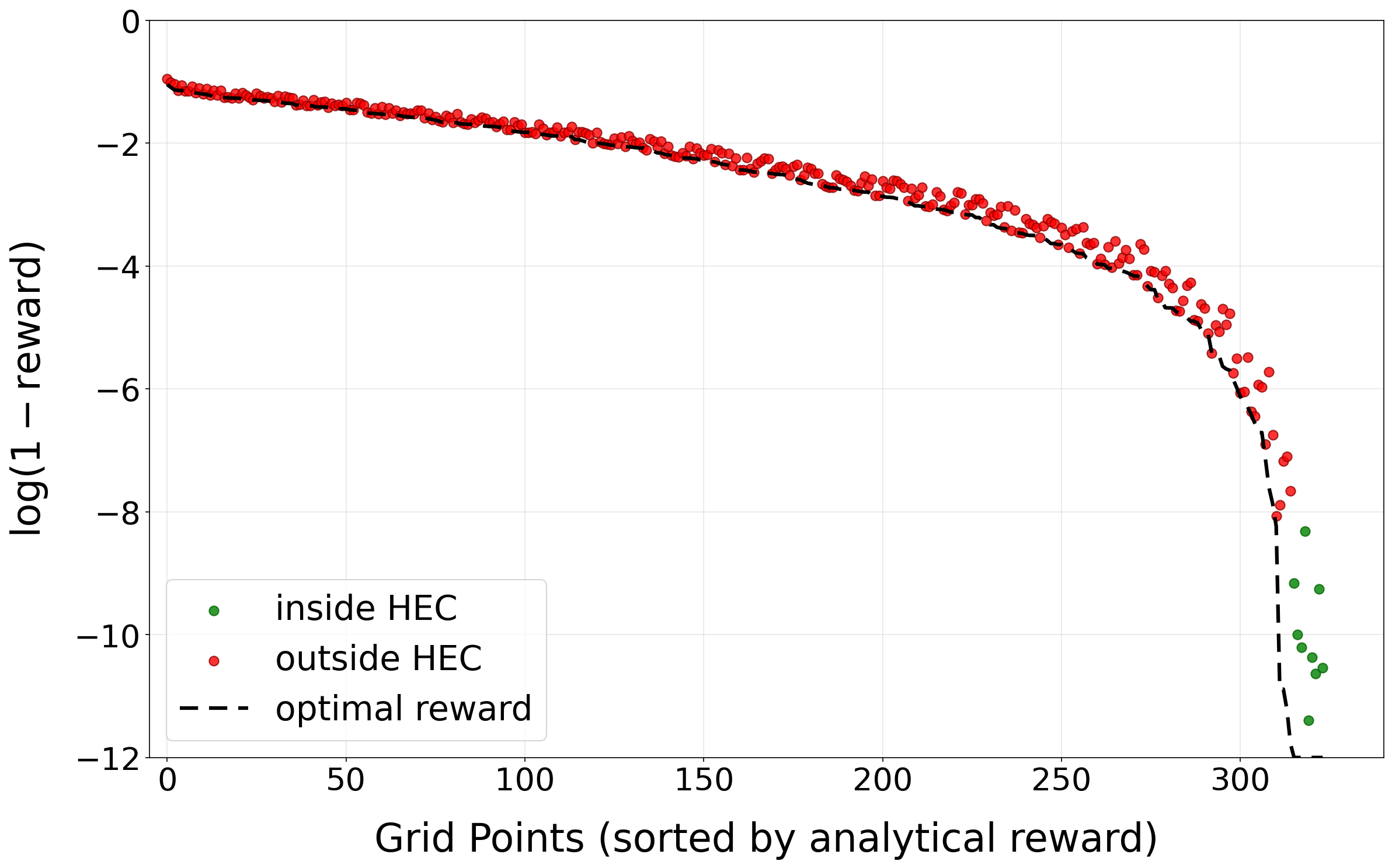}
    \caption{Classification comparison on the $\N=3$ symmetric slice with points sorted by analytical reward. Rewards are plotted as $\log(1-R)$ for visibility near the HEC boundary. Green points (inside HEC) cluster at high rewards near the analytical ceiling, while red points (outside HEC) follow the analytical curve with small gaps due to finite training.}
    \label{fig:rl-sorted}
\end{figure}

Beyond validating the reward landscape, we examine the classification power of the RL-achieved reward for distinguishing points inside versus outside the HEC. \Cref{fig:rl-sorted} presents a sorted classification view: all 324 grid points are arranged by analytical reward ($x$-axis) and compared against the maximum RL-achieved reward from 20 runs ($y$-axis).

The classification shows clean separation between inside and outside the HEC. All 9 points inside the HEC (green) achieve rewards very close to the theoretical maximum of 1, appearing as large negative values on the $\log(1-R)$ scale. The 315 points outside the HEC (red) achieve lower rewards that closely track the analytical predictions, with RL rewards consistently approaching the optimal dotted line from below. The gap between RL-achieved and analytical rewards reflects the finite training budget (20 runs, 2000 iterations each), yet remains small across the entire domain.

This clean separation demonstrates that the RL-achieved reward reliably distinguishes realizable from non-realizable entropy vectors. For points inside the HEC, the algorithm consistently finds near-optimal graph realizations; for points outside, it finds the best achievable realization on the HEC boundary. This classification capability provides the foundation for \Cref{sec:realizability}, where we apply the same approach to determine realizability of extreme rays in the $\N=6$ SAC.

\subsection{Gradient-based rediscovery of MMI}\label{ssec:mmi-rediscovery}

We will now demonstrate the effectiveness of the ``navigator'' function of the reward analyzed in \Cref{ssec:analytical-landscape}. In particular, we show that by starting from a vector outside the HEC, we can rediscover the MMI facet by simply following the reward gradient.

\paragraph{Experiment setup.}
We begin with the SAC extreme ray $\vec\ent_{\text{SAC}}$ given in \eqref{eq:sac-extreme-vec}, which lies in the MMI-violated region. The gradient is estimated using the RL algorithm with random perturbations orthogonal to $\vec\ent_{\text{SAC}}$: at each position, we sample 30 perturbed directions with perturbation sizes uniformly drawn from $[\delta S_{\max}/2, \delta S_{\max}]$ where $\delta S_{\max} = 0.02$, evaluate the reward at each perturbed point, and fit a linear model to extract the gradient direction (see Appendix~\ref{app:gradient-quality} for analysis of these parameters). Since the reward is scale-invariant, perturbations along the radial direction $\vec\ent$ provide no geometric information. We therefore sample perturbations in the $(2^\N-2)$-dimensional subspace orthogonal to $\vec\ent$, improving sample efficiency by focusing computational resources on geometrically meaningful directions.

At each iteration, we move a fixed step size of 0.1 in the estimated gradient direction, constrained to remain within the SA cone using the quadratic programming procedure described in Appendix~\ref{app:movement-algorithm}. This ensures that the trajectory respects all known SA constraints while progressing towards the unknown MMI boundary. We also include a momentum term (coefficient 0.3) that carries forward a fraction of the previous step direction, which helps smooth the trajectory but can cause overshoot when crossing boundaries.

\begin{figure}[tb]
    \centering
    \includegraphics[width=0.75\textwidth]{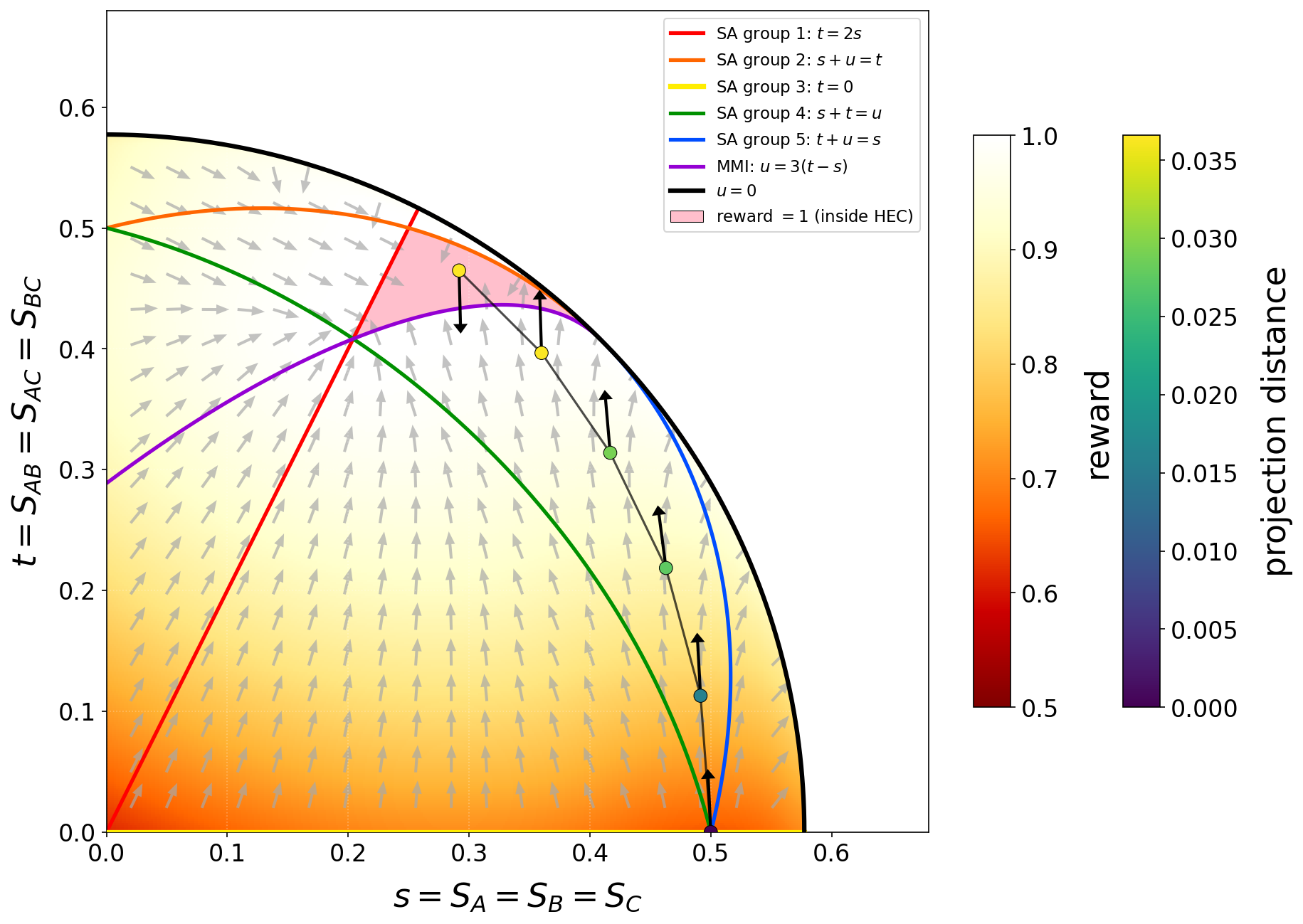}
    \caption{Trajectory of gradient-based optimization projected onto the symmetric slice $(s, t)$, shown until it first enters the HEC (pink region). The initial point lies exactly on the symmetric slice, and since the reward is higher on the symmetric slice than off it, the trajectory remains close to the slice throughout (projection distance $< 0.04$, shown by marker color). Black arrows indicate the gradient direction at each iteration; since we use fixed step size, all arrows are drawn with equal length. The 7D gradient $\vec{G}$ is projected to 2D via $\partial_s = G_A + G_B + G_C - 3s G_{ABC}/u$ and $\partial_t = G_{AB} + G_{AC} + G_{BC} - 3t G_{ABC}/u$, accounting for how $u$ varies with $s$ and $t$ on the unit sphere. These projected gradients appear to be not tangent to the trajectory due to the nonlinearity of the projection, but they align well with the analytical gradient field (gray arrows), confirming consistency between RL sampling and theoretical predictions. Detailed gradient evolution analysis is provided in~\Cref{fig:mmi-evolution}.}
    \label{fig:mmi-trajectory}
\end{figure}

\paragraph{Results.}
\Cref{fig:mmi-trajectory} shows the trajectory of the entropy vector projected onto the symmetric slice $(s, t)$, overlaid on the analytical reward landscape and its gradient direction. Starting from the SAC extreme ray $\vec\ent_{\text{SAC}}$ given in \eqref{eq:sac-extreme-vec} (which lies outside the HEC due to MMI violation), the trajectory moves steadily towards the MMI boundary while remaining inside the SAC. After approximately 5 iterations, the trajectory crosses into the HEC (pink region), at which point the reward nearly reaches its maximum value of 1.

\begin{figure}[tb]
    \centering
    \includegraphics[width=0.8\textwidth]{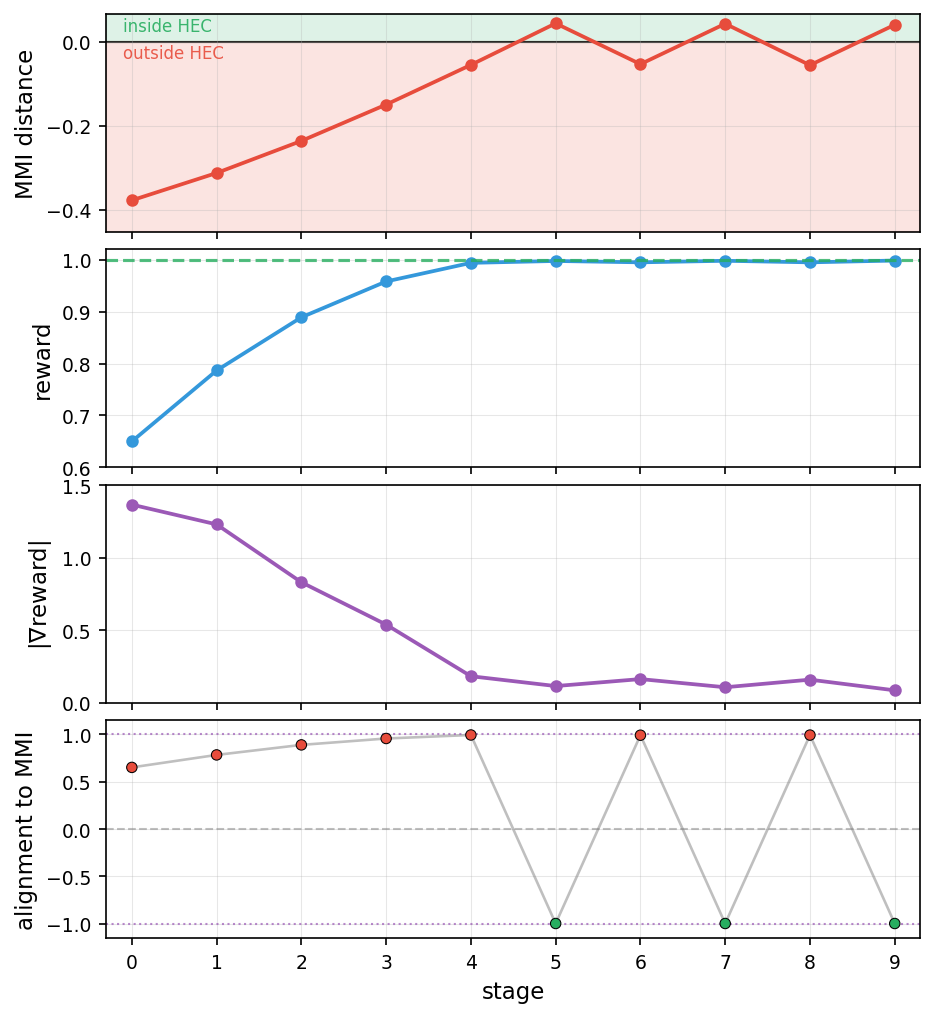}
    \caption{Evolution of key quantities during gradient-based movement towards the MMI boundary (stages 0--9). \textbf{Top}: Distance to MMI hyperplane, with green/red shading indicating inside/outside the HEC. The trajectory oscillates across the boundary due to fixed step size and momentum (see text). \textbf{Second}: Reward increases from 0.65 towards 1.0 as the trajectory approaches and enters the HEC. \textbf{Third}: Gradient magnitude $|\vec\nabla R|$ decreases near the HEC boundary where the reward landscape flattens. \textbf{Bottom}: Alignment (inner product) between the gradient and MMI facet normal. When outside the HEC (red points), the gradient points towards the MMI boundary; when inside (green points), it points away.}
    \label{fig:mmi-evolution}
\end{figure}

\Cref{fig:mmi-evolution} tracks four key quantities over 10 iterations (or stages) of optimization. The MMI distance (top panel) shows the trajectory starting outside the HEC (negative values) and crossing into the cone around stage 5, then oscillating back and forth across the boundary due to momentum overshoot. The reward (second panel) increases from approximately 0.65 towards 1 as the trajectory approaches and enters the HEC. The gradient magnitude (third panel) decreases as we approach the boundary, reflecting the flattening of the reward landscape inside the cone. Most importantly, the alignment between the gradient direction and the MMI facet normal (bottom panel) reveals how the gradient ``discovers'' the unknown facet.

\paragraph{Oscillation behavior.}
Theoretically, the reward equals unity everywhere inside the HEC, implying a vanishing gradient. However, as seen in \Cref{fig:mmi-evolution}, the RL-estimated gradient remains nonzero even inside the HEC, pointing back towards the MMI boundary. This apparent gradient likely arises because the RL optimization finds graph realizations more easily near the MMI boundary. Consequently, even though the true reward is unity throughout the interior, the \emph{estimated} reward from finite RL training is slightly higher near the boundary. If we had used a step size proportional to the gradient magnitude, the trajectory would saturate inside the HEC or exhibit only tiny oscillations as the gradient diminishes. Instead, our experiment employs a fixed step size, which amplifies the oscillation: the trajectory overshoots the boundary, detects a gradient pointing back, and reverses direction with the same step magnitude.

\paragraph{Implications.}
This experiment demonstrates three key capabilities of the RL-based approach. First, by constraining the trajectory to remain inside all \emph{known} facets (here, the SAs), we can flow towards \emph{unknown} facets and eventually enter the HEC. Second, the reward approaching unity and the gradient magnitude approaching zero serve as diagnostic indicators that the trajectory has reached or entered the HEC boundary, meaning we can detect proximity to the cone without prior knowledge of which facet we are approaching. Third, and most importantly, the gradient direction near the boundary reveals the normal vector of the constraining facet. As we approach the MMI facet, the sampled gradient aligns nearly perfectly with the vector normal to the MMI facet, allowing us to \emph{recover} the unknown inequality from the gradient alone. In this $\N=3$ demonstration, we treated the MMI inequality as unknown and successfully recovered it using only the SA constraints as prior knowledge. In the next section, we turn to the $\N=6$ HEC, where analytical treatment is no longer feasible and RL becomes essential for exploring the cone structure.

\section{Proving realizability of SAC extreme rays for $\N=6$}\label{sec:realizability}

We now turn to the $\N=6$ HEC, where both analytical treatment and traditional combinatorial search methods become intractable, making reinforcement learning an essential tool for exploring the cone structure. In particular, we will in this section demonstrate how we can use our RL algorithm to determine whether certain extreme rays of the $\N=6$ SAC are realizable and lie inside the $\N=6$ HEC.

\subsection{The 6 mystery rays}\label{ssec:mystery-rays}

In \cite{He:2024xzq}, the authors study the $\N=6$ SAC and determined its extreme rays. Excluding extreme rays that are uplifts of the SAC for fewer number of parties, there are 208 orbits of genuinely new extreme rays satisfying SSA, where each orbit is labeled by a representative ray, and all other extreme rays in that orbit can be obtained via permutation and purification symmetry. As expected, many of these extreme rays are not extreme rays of the HEC since they violate one of the known HEIs. However, there are 156 orbits that are candidates for extreme rays of the HEC, as they do not violate any known HEI. The graph realizations of 150 of these 156 orbits were found in \cite{He:2024xzq}, proving that they belong in the HEC and are extreme rays of the $\N=6$ HEC. However, \cite{He:2024xzq} could not find any graph realizations of the 6 remaining rays and dubbed them ``mystery rays.'' We now focus on these 6 mystery rays, which are given in \Cref{tab:mystery}.

\begin{table}[h]
\centering
\begin{tabular}{|c|c|}
\hline
{\small ER\#} & {\small $\vec \ent$ components} \\
\hline
\textbf{110} & \tiny{\{1,1,2,2,2,2; 2,3,3,3,3,3,3,3,3,4,4,4,4,4,4; 4,4,4,4,5,5,5,5,5,5,5,5,5,5,5,5,4,4,6,6; 6,4,4,6,6,6,5,5,5,5,5,5,5,5,4; 4,4,4,4,3,3; 2\}} \\
\hline
\textbf{145} & \tiny{\{2,2,2,3,3,3; 4,4,5,5,5,4,5,5,5,5,5,5,6,6,6; 6,7,7,7,7,7,7,6,6,8,7,7,7,8,8,6,8,8,8,9; 9,9,9,8,8,8,8,8,8,7,6,6,8,7,7; 6,6,6,5,5,5; 3\}} \\
\hline
\textbf{146} & \tiny{\{2,2,2,3,3,3; 4,4,5,5,5,4,5,5,5,5,5,5,6,6,6; 6,7,7,7,7,7,7,6,6,8,7,7,7,8,8,6,8,8,8,9; 9,9,9,8,8,8,8,8,6,7,6,8,8,7,7; 6,6,6,5,5,5; 3\}} \\
\hline
\textbf{168} & \tiny{\{2,2,2,3,3,3; 4,4,5,5,5,4,5,5,5,5,5,5,6,6,6; 6,7,7,7,7,7,7,6,6,8,7,7,7,8,8,6,8,8,8,9; 9,9,9,8,8,8,8,8,6,7,6,6,8,7,7; 6,6,6,5,5,5; 3\}} \\
\hline
\textbf{180} & \tiny{\{2,2,2,3,3,3; 4,4,5,5,5,4,5,5,5,5,5,5,6,6,6; 6,7,7,7,7,7,7,6,6,8,7,7,7,8,8,6,8,8,8,9; 9,9,9,8,8,8,6,8,6,7,8,6,8,7,7; 6,6,6,5,5,5; 3\}} \\
\hline
\textbf{181} & \tiny{\{2,2,2,3,3,3; 4,4,5,5,5,4,5,5,5,5,5,5,6,6,6; 6,7,7,7,7,7,7,6,6,8,7,7,7,8,8,6,8,8,8,9; 9,9,9,6,8,8,8,8,6,7,8,6,8,7,7; 6,6,6,5,5,5; 3\}} \\
\hline
\end{tabular}
\caption{We list the 6 mystery rays corresponding to the green rows from Table 4 of \cite{He:2024xzq}, keeping the original extreme ray (ER) numbering. The entropy vector components are arranged first by cardinality of the subsystem, and then in lexicographic order.}
\label{tab:mystery}
\end{table}

\subsection{RL training setup}\label{ssec:n6-setup}

We want to use our RL algorithm to determine whether any of these extreme rays have a graph realization. If there exists such graph realization, it must have 7 boundary vertices, which correspond to the 6 parties plus a purifier, as well as $n$ internal vertices.

For the $\N=6$ classification, we vary $n$ from 2 to 13 internal vertices, corresponding to total vertex counts $n_T =9$ to $n_T =20$. The policy network is a 4-layer feedforward network with 128 hidden units per layer. Training uses a batch size of 120, learning rate $10^{-4}$, rollout length of 100 steps, and a maximum of 3000 iterations per run. Early stopping with patience 7 is enabled. For each extreme ray and each value of $n_T$, we perform 20 independent training runs. For $n_T=9$, $12$, and $15$, we run all 208 extreme rays; for other values of $n_T$, we focus on the 6 mystery rays and selected candidates. To obtain robust statistics, we pool results across all $n_T$ values, yielding a total of over 7600 individual runs across the 208 extreme rays.

By designating each of the extreme rays in \Cref{tab:mystery} as a target vector $\vec\ent_{\rm target}$, we run the algorithm such that it attempts to find a graph with $n$ internal vertices whose min cuts reproduces $\vec\ent_{\rm target}$. Naturally, there are numerical errors associated to such a search, and from our RL training, we can only extract a list of edge weights corresponding to a graph whose min cuts approximate $\vec\ent_{\rm target}$. Importantly, however, once we have a candidate graph, it is easy to rescale the weights to integer values and check if the resultant graph has min cuts that is \emph{exactly} $\vec\ent_{\rm target}$.\footnote{For this task, we use a \texttt{Mathematica} script written by Massimiliano Rota as an independent check of our \texttt{Python} code.}

\subsection{Graph realizations found}\label{ssec:realizations}

\begin{figure}[tb]
    \centering
    \includegraphics[width=0.97\textwidth]{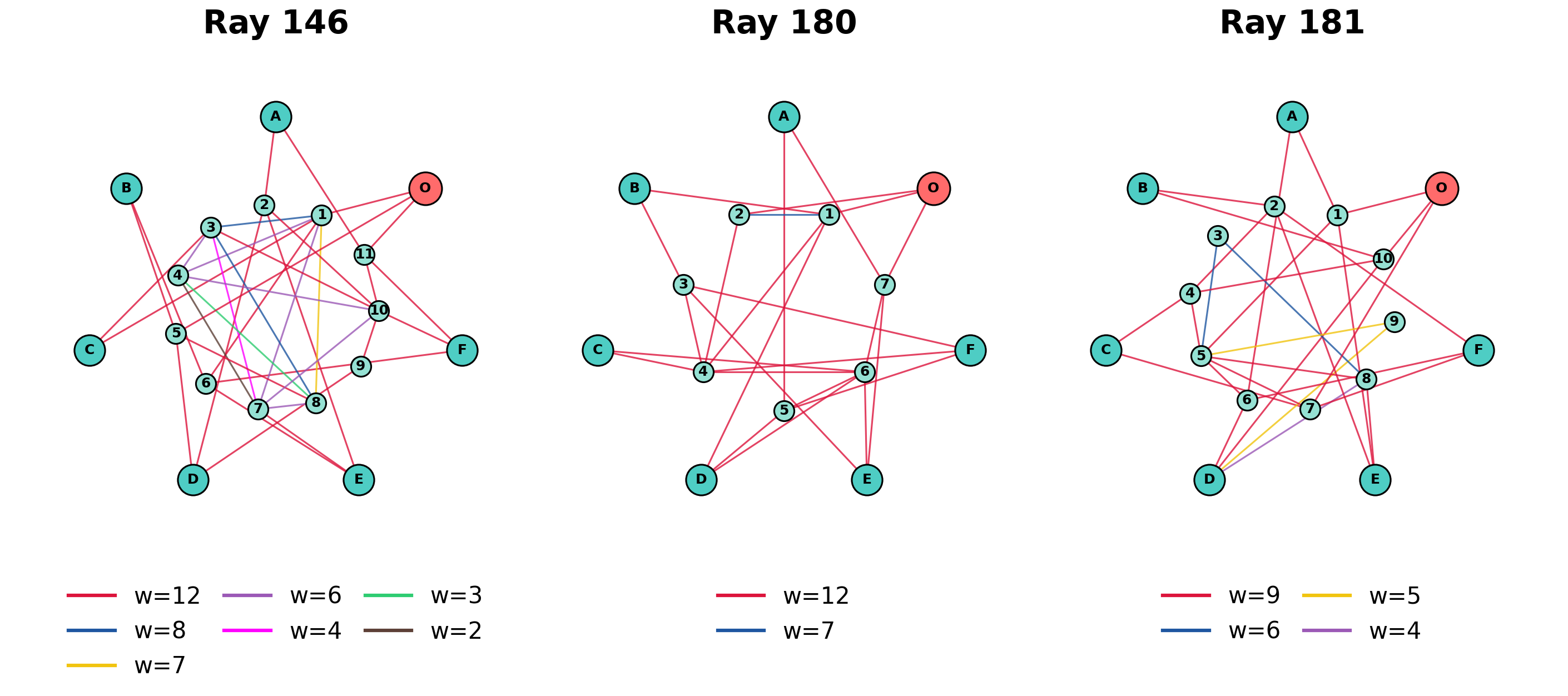}
    \caption{The graph realizations of extreme rays 146, 180, and 181.}
    \label{fig:graph_realizations}
\end{figure}

Applying this procedure to extreme rays 180 and 181, we are able to obtain in each case a weight vector associated to a graph whose edge weights are easily rescaled to be integers. The resultant graphs for extreme rays 180 and 181 are shown in \Cref{fig:graph_realizations}. Furthermore, for extreme ray 146, the RL algorithm also outputs a graph whose associated entropy vector is extremely close to the target entropy vector. However, it was more difficult to immediately guess how to rescale the edge weights to be integers that exactly reproduces extreme ray 146. Nevertheless, it was clear from the RL output how many internal vertices we need, which edges are present, and which edges have the same weight. Using these data, it was relatively straightforward to use trial and error to obtain a graph whose edge weights are integers and exactly reproduces extreme ray 146. The resultant graph is also shown in \Cref{fig:graph_realizations}. Because the graphs are rather complicated, we also for clarity represent each graph as a collection of weighted edges in the form $\{(v_1,v_2);w)\}$, where $(v_1,v_2)$ labels the two vertices the edge connects and $w$ its weight, below:
\begin{align*}
    \text{Ray 146:} \quad & \Big\{ \{(A,2);12\} , \{(A, 11);12\} , \{(B,5);12\} , \{(B,6);12\} , \{(C,1);12\} , \\
    &\qquad \{(C,3);12\} , \{(D,2);12\} , \{(D,5);12\} , \{(D,9);12\} , \{(E,2);12\} , \\
    &\qquad \{(E,6);12\} , \{(E,7);12\} , \{(F,6);12\} , \{(F,10);12\} , \{(F,11);12\} , \\
    &\qquad \{(O,1);12\} , \{(O,5);12\} , \{(O,11);12\} , \{(1,3);8\} , \{(1,4);6\} , \{(1,6);12\} , \\
    &\qquad \{(1,7);6\} , \{(1,8);7\} , \{(2,10);12\} , \{(3,4);6\} , \{(3,7);4\} , \{(3,8);8\} , \\
    &\qquad \{(3,10);12\} , \{(4,7);2\} , \{(4,8);3\} , \{(4,10);6\} , \{(5,8);12\} , \{(7,8);6\} , \\
    &\qquad \{(7,10);6\} , \{(9,10);12\} , \{(10,11);12\} \Big\}  \\
    \text{Ray 180:} \quad & \Big\{ \{(A,5);12\} , \{(A, 7);12\} , \{(B,1);12\} , \{(B,3);12\} , \{(C,4);12\} , \{(C,6);12\} , \\
    &\qquad \{(D,1);12\} , \{(D,5);12\} , \{(D,6);12\} , \{(E,3);12\} , \{(E,6);12\} , \\
    &\qquad \{(E,7);12\} , \{(F,3);12\} , \{(F,4);12\} , \{(F,5);12\} , \{(O,1);12\} , \\
    &\qquad \{(O,2);12\} , \{(O,7);12\} , \{(1,2);7\} , \{(1,4);12\} , \{(2,4);12\} , \{(3,4);12\} , \\
    &\qquad \{(4,6);12\} , \{(5,6);12\} , \{(6,7);12\} \Big\} \\
    \text{Ray 181:} \quad & \Big\{ \{(A,1);9\} , \{(A,6);9\} ,\{(B,2);9\} ,\{(B,10);9\} ,\{(C,4);9\} ,\{(C,7);9\} , \\
    &\qquad \{(D,6);9\} , \{(D,8);4\} , \{(D, 9);5\} , \{(D,10);9\} , \{(E,1);9\} , \{(E,2);9\} , \\
    &\qquad \{(E,8);9\} , \{(F,2);9\} , \{(F,6);9\} , \{(F,7);9\} , \{(O,1);9\} , \{(O,7);9\} , \\
    &\qquad \{(O,10);9\} , \{(1,5);9\} , \{(2,4);9\} , \{(3,5);6\} , \{(3,8);6\} , \{(4,5);9\} , \\
    &\qquad \{(4,10);9\} , \{(5,6);9\} , \{(5,7);9\} , \{(5,8);9\} , \{(5,9);5\} \Big\} .
\end{align*}

\subsection{Possible classification of all 208 rays}\label{ssec:classification}

Finally, we address the remaining 3 extreme rays 110, 145, and 168. The drawback of our algorithm is that because we are confined to a specific number of graph vertices, we cannot practically conclude with certainty that an extreme ray is not realizable by a graph, as we can always add additional internal vertices.\footnote{There is an upper bound on the total number of vertices necessary for any fixed $\N$, but this limit for $\N=6$ is 1,422,564 and is not practical \cite{Avis:2021xnz}.} Therefore, even though we have been unable to find a graph with $n \leq 13$ internal vertices, it is possible that if we include more internal vertices, we could potentially find a graph realization.

\begin{figure}
    \centering
    \includegraphics[width=1\linewidth]{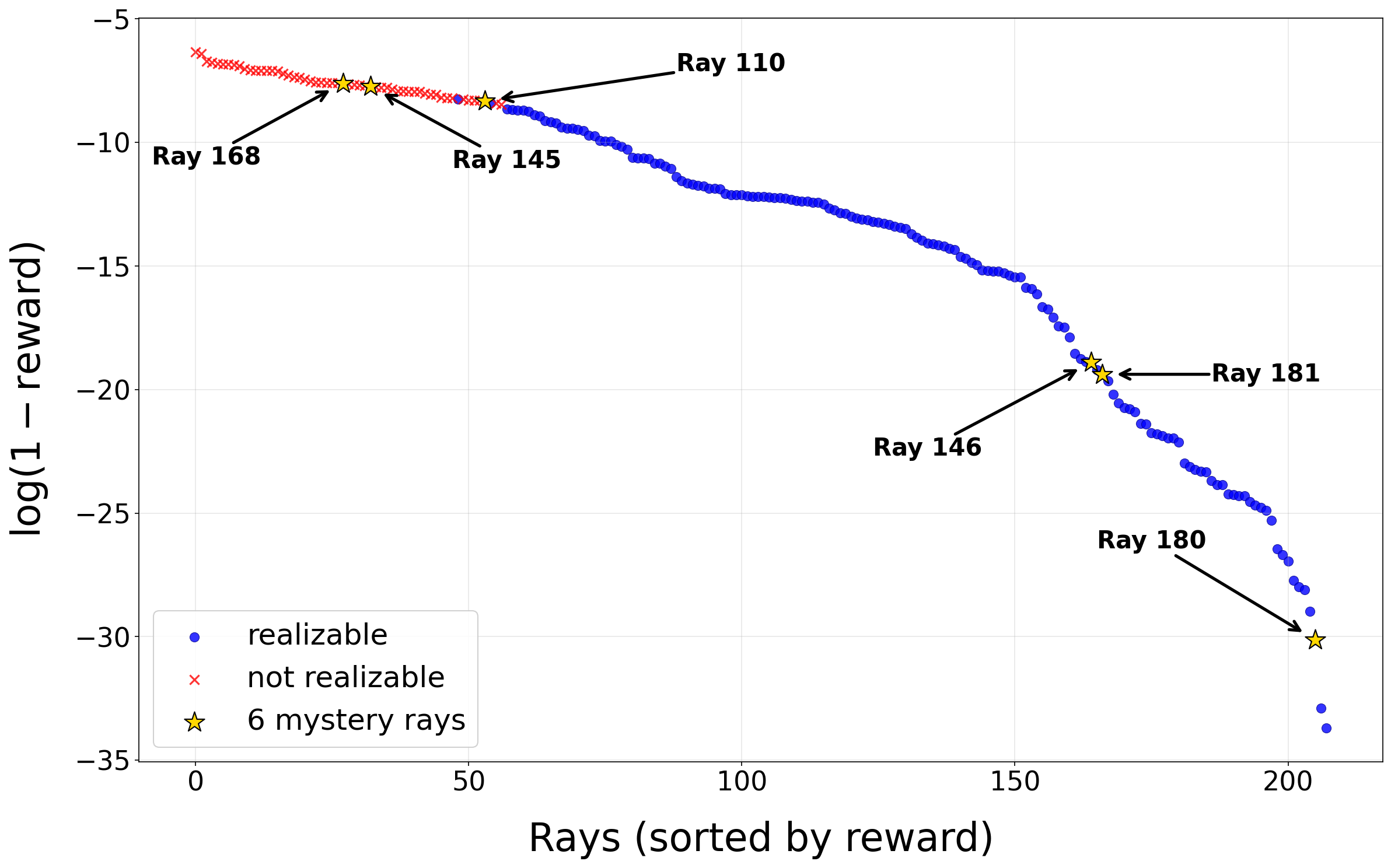}
    \caption{The maximum reward (pooled across multiple runs and $n_T$ values) associated to all 208 representatives of the $\N=6$ SAC extreme ray orbits. The red crosses denote extreme rays that are not realizable, while the blue dots denote the realizable extreme rays. The stars denote the 6 mystery extreme rays. Notice that there are 3 stars that are surrounded by blue dots, and these are precisely the extreme rays 146, 180, and 181 that we found realizations for in this paper. The remaining 3 mystery extreme rays are surrounded by red crosses, leading us to suspect they may not be genuine extreme rays of the $\N=6$ HEC.}
    \label{fig:208_class}
\end{figure}

Nevertheless, we now provide some evidence suggesting that the remaining 3 mystery extreme rays are actually not realizable. Again using our RL algorithm, except this time applying it to all 208 extreme rays found in \cite{He:2024xzq}, we plot the final reward output for each of these rays. The result is shown in \Cref{fig:208_class}. In this graph, the red crosses are associated to the 52 non-holographic extreme rays as they each violate a HEI. The blue dots represent the 150 holographic extreme rays, whose graph realizations were found in \cite{He:2024xzq}. Finally, the 6 stars the mystery extreme rays that we have been focused on in this paper. As is clear, 3 of the 6 mystery extreme rays are completely surrounded by holographic extreme rays, and these are precisely the extreme rays 146, 180, and 181 that we found graph realizations for in this paper (see \Cref{fig:graph_realizations}). However, the remaining 3 mystery rays are surrounded by extreme rays that violate some HEI, and therefore constitute evidence that they may perhaps also be non-holographic. If this is indeed the case, this means there must exist additional HEIs that were not discovered in \cite{Hernandez-Cuenca:2023iqh}, since these 3 SAC extreme rays must violate some HEI. Of course, this is not by any means a proof that the 3 remaining mystery rays are not realizable, and we leave a conclusive classification of them for future work.

\section{Discussion}\label{sec:discussion}

In this paper, we construct a simple RL algorithm to explore properties of the HEC. The purpose of the algorithm is to determine whether a particular target entropy vector can be associated to the min cuts of a graph with a fixed number of boundary and internal vertices. As a proof of principle, we applied the algorithm to an extreme ray of the SAC that lies outside the HEC due to MMI violation. Beginning with this non-holographic entropy vector, we move towards the HEC facet in entropy space through a series of iterations that increase the reward function. At a certain point, there will be a phase transition in the reward function, as we move from outside the HEC to inside the HEC. By verifying that the gradient of the reward function is orthogonal to the facet, we are able to determine that the facet corresponding to the phase transition of the reward function is precisely MMI.

Next, we shift our focus to the $\N=6$ HEC. The power of our RL algorithm is its ability to determine a graph realization for entropy vectors that are holographic. We thus apply it to the extreme rays of the $\N=6$ SAC that are compatible with SSA, which were determined in \cite{He:2024xzq}. There are 208 such orbits of extreme rays, and while \cite{He:2024xzq} was able to ascertain that 150 of them are holographic and 52 of them are not, there are 6 orbits that remain uncertain. Using our algorithm, we determine in this paper that 3 of them are realized holographically, albeit via rather complicated graphs that were not found in \cite{He:2024xzq} (see \Cref{fig:graph_realizations}). As for the remaining 3 orbits, we provide evidence that they are in fact not realizable by any graph construction (see \Cref{fig:208_class}).

There are several natural directions to pursue that will allow our algorithm to be more sophisticated and computationally more powerful. For instance, in our construction of possible graph realizations, our algorithm initiates the search with complete graphs. However, as was observed in \cite{Bao:2015bfa}, there are many graphs that give rise to the exact same entropy vector. If we can modify our algorithm to efficiently eliminate this redundancy, we would be able to speed up our efficiency in constructing possible graphs associated to a target entropy vector. In addition, one can use a hybrid algorithm involving both classical optimization techniques as well as reinforcement learning, and this is an avenue of research we are pursuing at the moment.

Furthermore, we can restrict ourselves to determining whether an entropy vector can be realized by not just any graph, but a tree graph. It was originally observed in \cite{Hernandez-Cuenca:2022pst} that all extreme rays of the $\N=5$ HEC can be realized by such tree graphs, and \cite{Hernandez-Cuenca:2022pst} further conjectured that this is true for any $\N$. By restricting ourselves to applying our RL algorithm to tree graphs only, we can attempt to find whether the known extreme rays of HEC at $\N=6$ can all be realized by tree graphs. Indeed, of the 150 classes of extreme rays of the $\N=6$ SAC that have a graph realization found in \cite{He:2024xzq}, 148 of them have tree graph realizations. It would be instructive to determine if the remaining 2 extreme rays also have tree graph realizations, potentially complementing the efficient algorithm developed in \cite{Hubeny:2024fjn, Hubeny:2025bjo, Hubeny:2025hst} to find tree graph realizations of entropy vectors.\footnote{We thank Veronika Hubeny for pointing out to us that our 3 new graphs in \Cref{fig:graph_realizations} all have tree graph realizations.}

It is remarkable that even with a vanilla policy gradient algorithm, we are able to make reasonable progress in the discovery of new extreme rays. This suggests that more sophisticated RL approaches could yield substantial improvements, in particular for the case of facet discovery. A first natural step would be to replace the vanilla policy gradient with a more modern on-policy algorithm such as Proximal Policy Optimization (PPO) \cite{Schulman:2017PPO}. Such an upgrade is expected to improve sample efficiency and stability. A second avenue is to recast the per target entropy vector search as a goal-conditioned or multi-task RL problem \cite{Schaul:2015UVFA, Andrychowicz:2017HER}, in which a single policy model is trained across a distribution of target entropy vectors and then deployed to new targets either zero-shot or with a brief fine-tuning phase. Because our gradient-based facet discovery in \Cref{sec:mmi} requires hundreds of independent RL runs per gradient estimate, amortizing this cost across targets through such pre-training could be especially impactful at $\N=6$.

Finally, we would also like to increase our precision of determining facets in the manner described in \Cref{sec:mmi}. While our gradient-based facet rediscovery demonstrated excellent results for $\N=3$ in \Cref{sec:mmi}, extending this approach to $\N=6$ proved significantly more challenging. The dimension of the $\N=6$ entropy space is $D = 2^6 - 1 = 63$, whereas that for the $\N=3$ entropy space is only 7. This high dimensionality, combined with the complex structure of the HEC with over 1800 known facet classes, requires substantially more gradient samples to obtain reliable estimates. In addition, each RL training run takes considerably longer at $\N=6$, making it impractical to gather sufficient statistics for meaningful gradient estimation within the scope of this work. We leave a more sophisticated search for new $\N=6$ HEIs using RL for future work.

\acknowledgments
 
We would like to thank Sergio Hern\'{a}ndez-Cuenca for useful discussions. We also acknowledge the assistance of Claude AI (Sonnet 3.7, Sonnet 4, Opus 4, Opus 4.1, Sonnet 4.5, and Opus 4.5), developed by Anthropic, which greatly accelerated our ability to test various hypotheses and iterate on experimental designs. This AI-assisted approach enabled rapid exploration of the parameter space and facilitated structured investigation of gradient estimation techniques.

This work is supported in part by the Walter Burke Institute for Theoretical Physics and the Leinweber Forums for Theoretical Physics at Caltech and by the U.S. Department of Energy, Office of Science, Office of High Energy Physics, under Award Number DE-SC0011632.  
T.H. is also supported by the Heising-Simons Foundation “Observational Signatures of Quantum Gravity” collaboration grant 2021-2817. H.O. is also supported in part by the Simons Investigator Award (MP-SIP-00005259) and JSPS Grants-in-Aid for Scientific Research 23K03379.
His work was performed in part at the Kavli Institute for the Physics and Mathematics of the Universe at the University of Tokyo, which is supported by the World Premier International Research Center Initiative, MEXT,
Japan, at the Kavli Institute for Theoretical Physics (KITP) at the University of California, Santa Barbara, which is supported by NSF grant PHY-2309135, at the Aspen Center for Physics, which is supported by NSF
grant PHY-1607611, and at Pioneering Science Promotion Division of RIKEN.

\appendix

\section{Analytical derivations for the $\N=3$ symmetric case}\label{app:analytical-derivations}

This appendix provides rigorous derivations of the optimal projection formulas used in \Cref{sec:mmi}, including proofs of optimality using Lagrange multipliers and second-order conditions.

\subsection{Problem statement}\label{app:problem}

\paragraph{Given.} Target entropy vector $\vec\ent_{\text{target}} = (s, s, s, t, t, t, u)$ with $\|\vec\ent\|^2 = 3s^2 + 3t^2 + u^2 = 1$.

\paragraph{Find.} Optimal achievable $\vec\ent' = (s', s', s', t', t', t', u')$ that maximizes cosine similarity, namely
\begin{align}\label{eq:app-maximize}
    \max_{s', t', u'} \quad R = 3ss' + 3tt' + uu' ,
\end{align}
subject to:
\begin{enumerate}
    \item $3s'^2 + 3t'^2 + u'^2 = 1$ (unit sphere)
    \item $t' \leq 2s'$ (SA group 1)
    \item $t' \leq s'+u'$ (SA group 2)
    \item $u' \leq 3(t' - s')$ (MMI)
    \item $s', t', u' \geq 0$
    \item SA groups $3,4$, and 5 (though these constraints are redundant given above constraints 1--5)
\end{enumerate}

\subsection{Case 1: target inside HEC}\label{app:case1}

\paragraph{Condition.} Target $(s, t, u)$ satisfies all HEC inequalities: $t \leq 2s$, $t\leq s+u$, and $u \leq 3(t - s)$.

\paragraph{Claim.} The optimal solution is $\vec\ent' = \vec\ent_{\text{target}}$, achieving $R = 1$.

\paragraph{Proof.} By the Cauchy--Schwarz inequality, we have for unit vectors
\begin{align}\label{eq:cs-ineq}
    R = \vec\ent_{\text{target}} \cdot \vec\ent' \leq \|\vec\ent_{\text{target}}\| \cdot \|\vec\ent'\| = 1,
\end{align}
with equality if and only if $\vec\ent' = \vec\ent_{\text{target}}$. Since the target satisfies all constraints, it is in the feasible region, so the maximum is achieved at the target itself with $R = 1$. $\square$

\subsection{Case 2a: SA group 1 violated ($t > 2s$)}\label{app:case2}

\paragraph{Condition.} $t > 2s$ (SA group 1 inequality is active at optimum).

\paragraph{Active constraint.} $t' = 2s'$.

\paragraph{Justification.} When the target violates SA group 1 inequality, the unconstrained optimum would be the target itself. However, this point is infeasible. The gradient of the reward function $\vec\nabla R = (3s, 3t, u)$ points towards the target, and since the target has $t/s > 2$, this gradient pushes towards the infeasible region. The optimum must therefore lie on the SA group 1 boundary $t' = 2s'$.

\paragraph{Substitution.} With $t' = 2s'$, the sphere constraint becomes $15s'^2 + u'^2 = 1$.

\paragraph{Reduced problem.}
\begin{align}\label{eq:app-R-SA}
    \max_{s', u'} \quad R = (3s + 6t)s' + uu' \quad \text{subject to} \quad 15s'^2 + u'^2 = 1 \text{ and } u' \leq 3s'.
\end{align}

\paragraph{Lagrangian.}
\begin{align}
    \mathcal{L} = (3s + 6t)s' + uu' - \lambda(15s'^2 + u'^2 - 1).
\end{align}

\paragraph{First-order conditions.}
\begin{align}
    \frac{\partial \mathcal{L}}{\partial s'} = 3s + 6t - 30\lambda s' = 0 \quad &\implies \quad s' = \frac{s + 2t}{10\lambda} \\
    \frac{\partial \mathcal{L}}{\partial u'} = u - 2\lambda u' = 0 \quad &\implies \quad u' = \frac{u}{2\lambda}.
\end{align}

\paragraph{Solution.} Substituting into $15s'^2 + u'^2 = 1$ and solving for $\lambda$:
\begin{align}
    \lambda = \frac{\sqrt{15(s+2t)^2 + 25u^2}}{10}.
\end{align}
This yields:
\begin{align}\label{eq:app-SA-sol}
    s' = \frac{s + 2t}{\sqrt{15(s+2t)^2 + 25u^2}}, \quad t' = 2s', \quad u' = \frac{5u}{\sqrt{15(s+2t)^2 + 25u^2}}.
\end{align}

\paragraph{Second-order condition.} The bordered Hessian analysis confirms this is a maximum since the determinant is positive for $\lambda > 0$ and $s', u' > 0$.

\paragraph{Feasibility check.} The interior solution requires $u' \leq 3s'$, which gives $5u \leq 3(s + 2t)$. If $5u > 3(s + 2t)$, the optimum is at the corner where both SA group 1 and MMI bind, namely $u' = 3s'$. This in turn implies on the unit sphere $24s'^2 = 1$, resulting in 
\begin{align}\label{eq:corner-solution}
    s' = \frac{1}{2\sqrt{6}}, \quad t' = \frac{1}{\sqrt{6}}, \quad u' = \frac{3}{2\sqrt{6}}.
\end{align}

\subsection{Case 2b: SA group 2 violated ($t > s+u$, with $t \leq 2s$)}\label{app:case2b}

\paragraph{Condition.} $t > s + u$ (SA group 2 inequality violated), but $t \leq 2s$ (SA group 1 satisfied).

\paragraph{Active constraint.} $t' = s' + u'$.

\paragraph{Justification.} When the target violates SA group 2 inequality $t \leq s + u$, the gradient of the reward $\vec\nabla R = (3s, 3t, u)$ points towards the target. Since the target has $t > s + u$, this gradient pushes towards the infeasible region. The optimum must therefore lie on the SA group 2 boundary $t' = s' + u'$.

\paragraph{Substitution.} With $t' = s' + u'$, the sphere constraint becomes:
\begin{align}
    3s'^2 + 3(s' + u')^2 + u'^2 = 1 \quad\implies\quad 6s'^2 + 6s'u' + 4u'^2 = 1.
\end{align}

\paragraph{Reduced problem.}
\begin{align}\label{eq:app-R-SA2}
    \max_{s', u'} \quad R = (3s + 3t)s' + (3t + u)u' \quad \text{subject to} \quad 6s'^2 + 6s'u' + 4u'^2 = 1.
\end{align}

\paragraph{Lagrangian.}
\begin{align}
    \mathcal{L} = (3s + 3t)s' + (3t + u)u' - \alpha(6s'^2 + 6s'u' + 4 u'^2 - 1).
\end{align}

\paragraph{First-order conditions.}
\begin{align}
    \frac{\partial \mathcal{L}}{\partial s'} &= 3s + 3t - \alpha (12 s' + 6 u') = 0   \\
    \frac{\partial \mathcal{L}}{\partial u'} &= 3t+u - \alpha (6 s' + 8 u') = 0  ,
\end{align}
which implies
\begin{align}\label{s1+u1}
    s' &= \frac{4s+t-u}{10\alpha} , \qquad u' = \frac{3(t-s)+2u}{10\alpha} .
\end{align}

\paragraph{Solution.} It is useful to define auxiliary quantities 
\begin{align}
\begin{split}
    P &= 4s + t - u, \quad Q = 3(t-s) + 2u , \quad D = \sqrt{6P^2 + 6P Q  + 4Q^2}.
\end{split}
\end{align}
It follows upon solving for $\alpha$ given $6s'^2 + 6 s'u' + 4 u'^2 = 1$ given \eqref{s1+u1} that
\begin{align}\label{alpha}
    \alpha = \frac{D}{10}.
\end{align}
Substituting \eqref{alpha} into \eqref{s1+u1} and using the above auxiliary quantities, we find the optimal solution
\begin{align}\label{eq:app-SA2-sol}
    s' = \frac{P}{D}, \quad t' = \frac{P + Q}{D}, \quad u' = \frac{Q}{D}.
\end{align}

\paragraph{Second-order condition.} The bordered Hessian analysis confirms this is a maximum since the determinant is positive for the relevant parameter range.

\paragraph{Feasibility check.} The interior solution requires $t' \leq 2s'$, which translates to $Q \leq P$. If $Q > P$, the optimum is at the corner where both SA group 1 and SA group 2 bind, namely $t' = 2s'$ and $t' = s' + u'$, giving $u' = s'$. This in turn implies on the unit sphere $16s'^2=1$, resulting in
\begin{align}\label{eq:corner-SA2}
    s' = \frac{1}{4}, \quad t' = \frac{1}{2}, \quad u' = \frac{1}{4}.
\end{align}

\subsection{Case 3: MMI violated ($u > 3(t-s)$, with $t \leq 2s$)}\label{app:case3}

\paragraph{Condition.} $u > 3(t - s)$ and $t \leq 2s$.

\paragraph{Active constraint.} $u' = 3(t' - s')$.

\paragraph{Reduction to single variable.} Let $\rho = t'/s'$. Then $t' = \rho s'$ and $u' = 3(\rho - 1)s'$. The unit sphere constraint is now given by
\begin{align}
    s' = \frac{1}{\sqrt{6(2\rho^2 - 3\rho + 2)}}.
\end{align}

\paragraph{Optimization over $\rho$.} Setting $A = s - u$ and $B = t + u$, the reward \eqref{eq:app-maximize} becomes
\begin{align}\label{eq:app-R-MMI}
    R = \frac{3(A + B\rho)}{\sqrt{6(2\rho^2 - 3\rho + 2)}}.
\end{align}
Taking the derivative and setting it to zero yields
\begin{align}
    \rho = \frac{3s + 4t + u}{4s + 3t - u}.
\end{align}

\paragraph{Validity range.} The ratio $\rho$ must satisfy $1 \leq \rho \leq 2$. The lower bound follows from requiring $s',t',u'>0$, and the upper limit follows from SA group 1.
\begin{itemize}
    \item At $\rho = 1$: $u' = 0$, point on $t' = s'$ line.
    \item At $\rho = 2$: $t' = 2s'$, corner with SA group 1.
\end{itemize}
If computed $\rho < 1$, instead impose $\rho = 1$. If computed $\rho > 2$, instead impose $\rho=2$ and use the corner solution \eqref{eq:corner-solution}.

\paragraph{Final formula.}
\begin{align}\label{eq:optimal-rho-derivation}
\begin{split}
    \rho &= \frac{3s + 4t + u}{4s + 3t - u} \quad \text{(clamped to $\rho \in [1, 2]$)} \\
    s' &= \frac{1}{\sqrt{6(2\rho^2 - 3\rho + 2)}}, \quad t' = \rho s', \quad u' = 3(\rho - 1)s'.
\end{split}
\end{align}

\subsection{Explicit reward formulas}\label{app:reward-formulas}

\paragraph{SA group 1 projection reward.} For targets in the SA group 1 violated region, we have upon substituting \eqref{eq:app-SA-sol} into \eqref{eq:app-R-SA}
\begin{align}
    R_{\text{SA}_1} = \frac{3(s+2t)^2 + 5u^2}{\sqrt{15(s+2t)^2 + 25u^2}}.
\end{align}

\paragraph{SA group 2 projection reward.} For targets in the SA group 2 violated region, we have upon substituting \eqref{eq:app-SA2-sol} into \eqref{eq:app-R-SA2}
\begin{align}
    R_{\text{SA}_2} = \sqrt{\frac{2[6s^2+6t^2 + 3s(t-u)+3tu+u^2]}{5}} .
\end{align}

\paragraph{MMI projection reward.} For targets in the MMI-violated region, we have using \eqref{eq:app-R-MMI}
\begin{align}
    R_{\text{MMI}} = \frac{3[s + (t + u)\rho - u]}{\sqrt{6(2\rho^2 - 3\rho + 2)}},
\end{align}
where $\rho$ is given by \eqref{eq:optimal-rho-derivation}.

\subsection{Summary of projection formulas}\label{app:summary}

We summarize the optimal projection for each region of the $(s, t)$ plane:

\begin{center}
\begin{tabular}{|l|l|l|}
\hline
\textbf{Region} & \textbf{Condition} & \textbf{Optimal $(s', t', u')$} \\
\hline
Inside HEC & All satisfied & $(s, t, u)$ \\
\hline
SA group 1 violated & $t > 2s$, $5u \leq 3(s+2t)$ & $D_1 = \sqrt{15(s+2t)^2 + 25u^2}$ \\
(interior) & & $s' = \frac{s+2t}{D_1}$, $t' = 2s'$, $u' = \frac{5u}{D_1}$ \\
\hline
SA group 1 violated & $t > 2s$, $5u > 3(s+2t)$ & $s' = \frac{1}{2\sqrt{6}}$, $t' = \frac{1}{\sqrt{6}}$, $u' = \frac{3}{2\sqrt{6}}$ \\
(corner) & & \\
\hline
SA group 2 & $t > s+u$, $t \leq 2s$, & $P = 4s+t-u$, $Q = 3(t-s)+2u$, \\
violated (interior) & $Q \leq P$ & $D = \sqrt{6P^2 + 6PQ + 4Q^2}$, \\
& & $s' = \frac{P}{D}$, $t' = \frac{P+Q}{D}$, $u' = \frac{Q}{D}$ \\
\hline
SA group 2 & $t > s+u$, $t \leq 2s$, & $s' = \frac{1}{4}$, $t' = \frac{1}{2}$, $u' = \frac{1}{4}$ \\
violated (corner) & $Q > P$ & \\
\hline
MMI violated & $u > 3(t-s)$ or $u > 2t-s$, & $\rho = \frac{3s+4t+u}{4s+3t-u}$, $s' = \frac{1}{\sqrt{6(2\rho^2-3\rho+2)}}$, \\
& with $t \leq 2s$ & $t' = \rho s'$, $u' = 3(t'-s')$ \\
\hline
\end{tabular}
\end{center}

\noindent All formulas are proven optimal via Lagrange multipliers with verified second-order conditions. $\square$

\section{Gradient estimation quality}\label{app:gradient-quality}

Reliable gradient estimation is crucial for gradient-based facet discovery. In this appendix, we analyze how the perturbation size $\delta S$ and the number of gradient samples affect estimation quality by using the $\N=3$ symmetric case as a benchmark. All experiments in this appendix are conducted at the entropy vector $\vec\ent_{\text{GHZ}} := \frac{1}{7}\{1,1,1;1,1,1;1\}$ corresponding to the GHZ state, which lies outside the HEC and provides a well-defined analytical gradient for comparison.

\subsection{Signal-to-noise ratio and perturbation size}\label{app:snr}

Our gradient estimation method uses random orthogonal perturbations followed by linear regression to extract the gradient direction. For each perturbation direction, the perturbation magnitude is drawn uniformly from $[\delta S_{\max}/2, \delta S_{\max}]$, which provides a range of scales for robust linear fitting. The quality of this estimate depends critically on the signal-to-noise ratio (SNR): the gradient signal must be strong enough relative to the noise in the reward measurements.

When $\delta S$ is too small, the reward change $\Delta R$ induced by the perturbation approaches the numerical precision limit, leading to gradient estimates dominated by noise. Conversely, when $\delta S$ is too large, the linear approximation $R(\vec\ent + \delta\vec\ent) \approx R(\vec\ent) + \vec\nabla R \cdot \delta\vec\ent$ breaks down, as the reward landscape has curvature that invalidates linear regression.

\begin{figure}[t]
    \centering
    \includegraphics[width=0.9\textwidth]{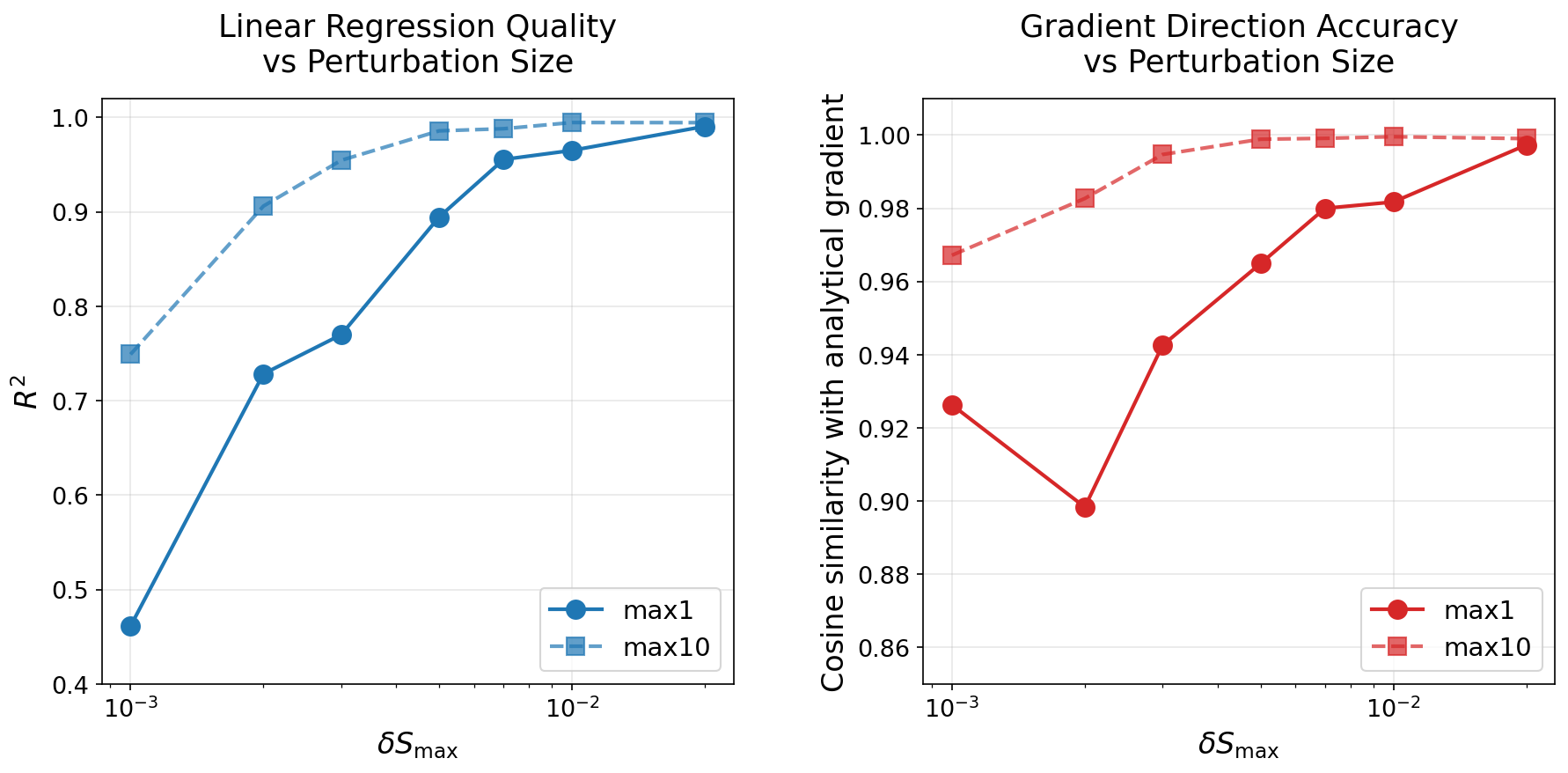}
    \caption{Gradient estimation quality as a function of perturbation size $\delta S$, comparing single-sample ({\sf max1}) and 10-sample pooling ({\sf max10}) strategies. \textbf{Left:} Coefficient of determination $\mathbf R^2$ of the linear regression. \textbf{Right:} Cosine similarity between the estimated gradient and the analytically computed gradient for the symmetric test point. Both metrics show that {\sf max10} pooling improves quality at low $\delta S$ by reducing noise, but both strategies converge for $\delta S \gtrsim 0.01$. Quality peaks near $\delta S \approx 0.02$.}
    \label{fig:dS-gradient-quality}
\end{figure}

\Cref{fig:dS-gradient-quality} shows both the coefficient of determination $\mathbf{R}^2$ of the linear regression and the cosine similarity with the analytically computed gradient, as a function of $\delta S$. We compare single-sample ({\sf max1}) and multi-sample ({\sf max10}) pooling strategies. Several observations emerge:
\begin{enumerate}
    \item For very small $\delta S < 0.005$, gradient quality degrades due to low SNR.
    \item The {\sf max10} pooling strategy (taking the maximum reward across 10 RL runs at each perturbed point) can partially compensate for low $\delta S$ by reducing measurement noise.
    \item For sufficiently large $\delta S \gtrsim 0.01$, {\sf max1} and {\sf max10} converge to similar quality, indicating that noise is no longer the limiting factor.
    \item Quality peaks around $\delta S \approx 0.02$ with $\mathbf R^2 > 0.99$ and cosine similarity $> 0.99$ with the analytical gradient, while it is not yet declined for larger $\delta S$ as the linear approximation breaks down.
\end{enumerate}

\noindent Based on this analysis, we chose $\delta S_{\max} = 0.02$ for our experiments, which achieves $\mathbf R^2 > 0.99$ and excellent directional accuracy ($> 99\%$ cosine similarity with the analytical gradient) while maintaining valid linear approximations.

\subsection{Sample number and stability}\label{app:samples}

The number of gradient samples $N_{\text{samples}}$ affects both the accuracy and stability of gradient estimates. More samples provide better coverage of the perturbation directions and more robust linear regression fits.

\begin{figure}[t]
    \centering
    \includegraphics[width=0.6\textwidth]{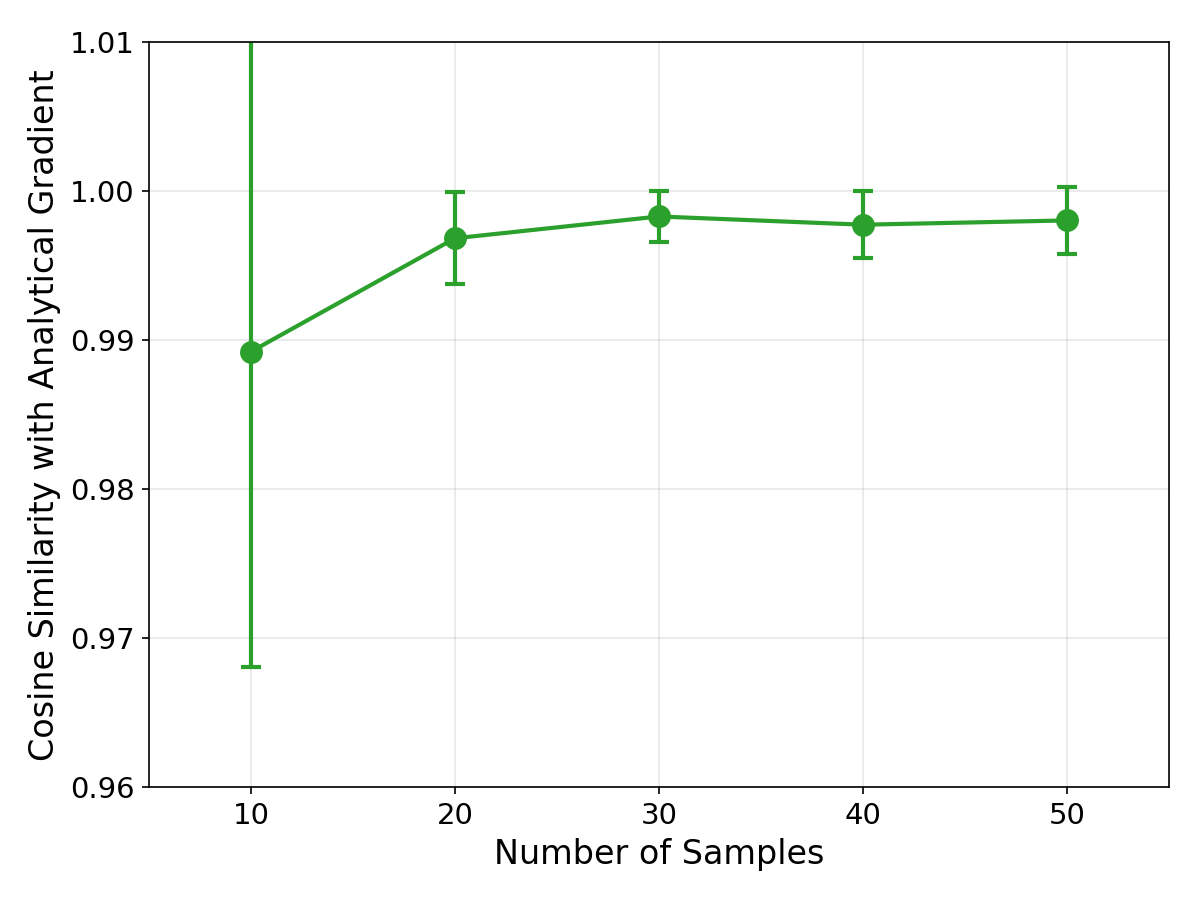}
    \caption{Cosine similarity between the estimated gradient and the analytically computed gradient as a function of sample number, with error bars showing 1 standard deviation variability across 10 independent runs. Alignment improves rapidly up to 30 samples and stabilizes beyond.}
    \label{fig:samples-gradient-quality}
\end{figure}

\Cref{fig:samples-gradient-quality} shows the cosine similarity between the estimated and analytical gradients as a function of sample number, using $\delta S_{\max} = 0.02$ and the {\sf max1} pooling strategy. Error bars indicate variability across 10 independent runs. We observe:
\begin{enumerate}
    \item Gradient quality improves rapidly up to approximately 30 samples.
    \item Beyond 30 samples, improvements begin to saturate, with diminishing returns.
    \item Stability (smaller error bars) also improves with more samples, which is important for reproducibility.
\end{enumerate}

\noindent Based on these observations, we chose 30--50 gradient samples for our experiments as the point where quality improvements begin to saturate while maintaining reasonable computational cost.

\section{Gradient-constrained movement algorithm}\label{app:movement-algorithm}

When moving towards unknown HEC facets from a point outside the cone, we must ensure that movement respects all \emph{known} facet constraints while exploring in the gradient direction. This appendix describes our constrained movement algorithm.

\subsection{Problem setup}\label{app:movement-setup}

Consider an entropy vector $\vec\ent$ that lies outside the HEC, violating some unknown facet. We have computed the reward gradient $\vec\nabla R$ pointing towards the HEC boundary. Our goal is to find a movement direction $\vec{d}$ such that:
\begin{enumerate}
    \item It is as close as possible to the gradient direction $\vec\nabla R$.
    \item It does not violate any \emph{known} HEC facets $\vec{a}_i \cdot \vec\ent \geq 0$.
\end{enumerate}

\noindent This is formulated as a quadratic programming (QP) problem:
\begin{align}\label{eq:qp-safe}
\begin{split}
    &\min_{\vec{d}} \, \|\vec{d} - \vec\nabla R\|^2 \\
    \text{such that} \quad & \vec{a}_i \cdot \vec{d} \geq 0 \quad \text{for all known facets } i \text{ with } \vec{a}_i \cdot \vec\ent < \epsilon_{\text{sat}}, 
\end{split}
\end{align}
where $\epsilon_{\text{sat}}$ is a small saturation tolerance (we use $10^{-4}$). The constraint is only imposed for facets that are nearly saturated, as facets far from saturation will not be violated by small movements.

\subsection{Safe distance computation}\label{app:safe-distance}

After finding the optimal direction $\vec{d}^*$, we compute the maximum safe distance we can move before violating any facet:
\begin{align}
    d_{\text{safe}} = \min_i \left\{ \frac{\vec{a}_i \cdot \vec\ent}{-\vec{a}_i \cdot \vec{d}^*} \;\Bigg|\; \vec{a}_i \cdot \vec{d}^* < 0 \right\}.
\end{align}
The actual movement distance is then $\min(d_{\text{safe}}, \vec v \cdot \vec\nabla R)$, where $\vec v$ is the velocity parameter.

\subsection{Escape mode for gradient contamination}\label{app:escape-mode}

A subtle issue arises when the current position is very close to known facet boundaries (within distance $\delta S$ used for gradient sampling). In this case, gradient samples may cross the known facet boundaries, leading to \emph{contaminated} gradient estimates that point towards the known facets rather than the unknown violated facet.

To address this, we introduce \emph{escape mode} with two mechanisms:
\begin{enumerate}
    \item \textbf{Buffer constraint}: Maintain a buffer zone of size $\beta \times \delta S$ (typically $\beta = 1.5$) from all known facets. This ensures gradient samples do not cross known boundaries.

    \item \textbf{Escape constraint}: For facets within the escape threshold $\theta_{\text{escape}} = 1.5 \times \beta \times \delta S$, we impose an additional constraint requiring active movement away from these facets:
    \begin{align}
        \vec{a}_i \cdot \vec{d} \geq r_{\text{min}} \quad \text{for facets with } \vec{a}_i \cdot \vec\ent < \theta_{\text{escape}},
    \end{align}
    where $r_{\text{min}}$ is the minimum increase rate (typically 0.05--0.15).
\end{enumerate}

This two-tier system ensures that:
\begin{itemize}
    \item Facets near the escape threshold are actively moved away from, preventing gradient contamination.
    \item Facets above the threshold are respected via distance constraints.
    \item The gradient signal comes primarily from the unknown violated facet.
\end{itemize}

\noindent When the escape-mode QP becomes infeasible (no direction satisfies all constraints), the algorithm automatically falls back to standard safe-direction mode without the escape constraints.

\bibliography{entropy-cone-bib}{}
\bibliographystyle{utphys}

\end{document}